\documentclass[]{tCPH2e}

\begin{document}

\jvol{00} \jnum{00} \jyear{0000} 

\markboth{T.N.Palmer}{Contemporary Physics}


\title{Lorenz, G\"{o}del and Penrose: New Perspectives on Determinism and Causality in Fundamental Physics}
\author{T.N.Palmer\thanks{Email: tim.palmer@physics.ox.ac.uk
\vspace{6pt}} \\\vspace{6pt}{\em{University of Oxford, Clarendon Laboratory, Parks Rd, Oxford}}\\\vspace{6pt}\received{v3.0 released March 2014} }

\makeatletter
\newcommand\be{\@ifstar{\[}{\begin{equation}}}
\newcommand\ee{\@ifstar{\]}{\end{equation}}}
\newcommand\bp{\begin{pmatrix}}
\newcommand\ep{\end{pmatrix}}
\newcommand\ua{\uparrow}
\newcommand\da{\downarrow}
\makeatother 

\bibliographystyle{plain}

\maketitle

\begin{abstract}
Despite being known for his pioneering work on chaotic unpredictability, the key discovery at the core of meteorologist Ed Lorenz's work is the link between space-time calculus and state-space fractal geometry. Indeed, properties of Lorenz's fractal invariant set relate space-time calculus to deep areas of mathematics such as G\"{o}del's Incompleteness Theorem. These properties, combined with some recent developments in theoretical and observational cosmology, motivate what is referred to as the `cosmological invariant set postulate': that the universe $U$ can be considered a deterministic dynamical system evolving on a causal measure-zero fractal invariant set $I_U$ in its state space. Symbolic representations of $I_U$ are constructed explicitly based on permutation representations of quaternions. The resulting `invariant set theory' provides some new perspectives on determinism and causality in fundamental physics. For example, whilst the cosmological invariant set appears to have a rich enough structure to allow a description of quantum probability, its measure-zero character ensures it is sparse enough to prevent invariant set theory being constrained by the Bell inequality (consistent with a partial violation of the so-called measurement independence postulate). The primacy of geometry as embodied in the proposed theory extends the principles underpinning general relativity. As a result, the physical basis for contemporary programmes which apply standard field quantisation to some putative gravitational lagrangian is questioned. Consistent with Penrose's suggestion of a deterministic but non-computable theory of fundamental physics, a  `gravitational theory of the quantum' is proposed based on the geometry of $I_U$, with potential observational consequences for the dark universe.
 \end{abstract}

\section{Introduction} 
\label{introduction}

\begin{figure}
\centering
\includegraphics{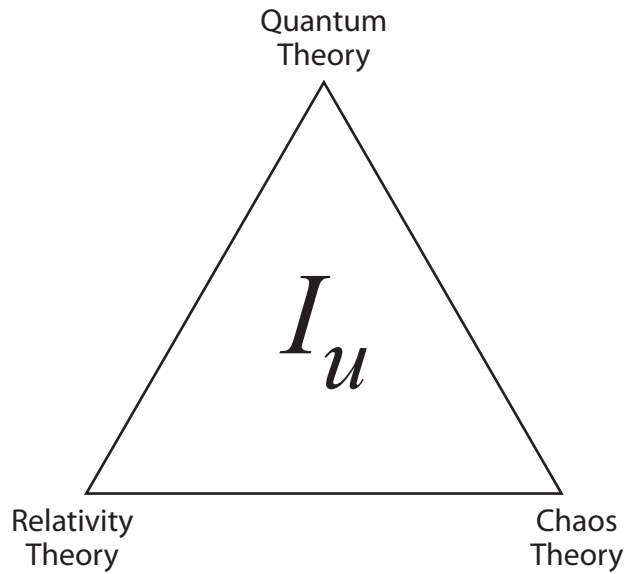}
\caption{It is proposed that the three great revolutions of 20th Century physics can be unified if the universe is considered a deterministic dynamical system evolving on a fractal invariant set $I_U$ in state space.}
\label{fig:triangle}
\end{figure}

There were three great revolutions in 20th Century theoretical physics: relativity theory, quantum theory and chaos theory (Fig \ref{fig:triangle}). Each has had a profound impact on the development of science, and yet their domains of impact remain quite distinct. Despite over a half century of intense research, there is still no consensus on how to combine quantum theory and general relativity theory into a supposed `quantum theory of gravity', nor even a consensus about what such a notion means physically (left edge of triangle). Moreover, the unpredictability of nonlinear chaotic systems is generally considered quite unrelated to the indeterminism of quantum measurement (right edge of triangle). Finally, whilst an essential characteristic of chaos is the existence of positive Lyapunov exponents (measuring exponential divergence of neighbouring state-space trajectories in time) this is not a relativistically invariant characteristic; under a logarithmic transformation of time, a positive exponent can be transformed to a zero exponent (bottom edge of triangle). The purpose of this paper is to develop new ideas which may lead to some unification of these three revolutions. These ideas evolve around Einstein's great insight that geometry provides the ultimate expression of the laws of physics. Motivated by developments in cosmology on the one hand and nonlinear dynamics on the other, here we attempt to extend this geometric insight from space-time to state-space.  The key idea which motivates the discussion in this paper is the fractal state-space geometry associated with certain classes of nonlinear dynamical system exemplified by the Lorenz equations \cite{Lorenz:1963} (see Section \ref{remarkable}). The key cosmological developments which justify this discussion are those of quasi-cyclic cosmologies, the discovery of a positive cosmological constant and the black-hole no-hair theorem. 

In Section \ref{remarkable} it is shown how properties of Lorenzian fractal geometry have links to one of the most famous theorem in 20th Century mathematics - the G\"{o}del incompleteness theorem - and indeed also to some of the number-theoretic tools used in Wiles' proof of Fermat's Last Theorem. If this geometry relates to deep concepts from 20th Century mathematics, could it also relate to deep but still troublesome concepts from 20th Century physics; concepts such as state superposition, incompatible observables and quantum non-locality? Above all, could a focus on state-space geometry provide some new perspectives on the problem of `quantum gravity'? In Section \ref{sec:liouville}, the strong formal similarities between the Schr\"{o}dinger equation and the classical Liouville equation for conservation of state-space probability are discussed. Not least, both equations are linear. Just as the linearity of the Liouville equation is no constraint to how nonlinear and chaotic may be the underpinning deterministic evolution equations, is it possible that the linearity of the Schr\"{o}dinger equation hides some underpinning deterministic dynamic that is also profoundly nonlinear? The standard view is that it is not possible if we require these dynamics to be locally causal; this is the Bell Theorem, a focus for discussion in this paper. 

By introducing a causal concept called `the cosmological invariant set postulate' \cite{Palmer:2009a}, an important conclusion of this paper is that this standard acausal interpretation of the Bell Theorem is incorrect. The cosmological invariant set postulate, that the universe $U$ is evolving causally and deterministically on some measure-zero fractal invariant set $I_U$ in its state space, is introduced in Section \ref{universe}. As discussed in Section \ref{belltheorem}, this postulate is a potential game changer as far as the Bell Theorem is concerned, raising the possibility that the experimentally verified existence of nonclassical correlations can be explained deterministically without resorting to any violation of local causality in space-time (the technical role of the invariant set postulate in the Bell Theorem is that it provides a non-conspiratorial cosmological basis for a partial violation of the so-called measurement independence postulate \cite{Hall:2010} \cite{Hall:2011}). In Section \ref{quantum} these ideas are developed more explicitly. In particular, a symbolic representation of $I_U$ is constructed; from this it is shown how key properties of qubit quantum physics are emergent. `Invariant set theory' provides the means to explain the three key differences between the Schr\"{o}dinger equation and the classical Liouville equation: the existence of the square root of minus one and of Planck's constant (Section \ref{complex}), and of the complex Hilbert Space (Section \ref{hilbert}). In particular the symbolic construction of $I_U$ makes use of a novel permutation/negation representation of complex numbers and it is outlined how the complex Hilbert Space of quantum theory can be viewed as the singular limit \cite{Berry} of symbolic bit strings associated with the construction. In Section \ref{emergent}, two features which emerge from this causal deterministic construction are discussed: the notion of incompatible observables and violation of the Bell inequalities. In both cases, the emergence of these features hinges on an application of simple (but profound) number-theoretic properties of the cosine function. As discussed, in Section \ref{emergent}, these features also allows a new perspective on the iconic two-slit experiment. 

This paper is based on the author's 9th Dennis Sciama Memorial Lecture given in 2013 \footnote{A video of the lecture is available at https://vimeo.com/61514195}. The author was a student of Dennis Sciama in the 1970s, working on the gravitational energy-momentum problem in general relativity theory \cite{Palmer:1978} \cite{Palmer:1980}. During this period, he was strongly influenced by the work of Roger Penrose, one of the protagonists in the title of this paper. As Penrose wrote in the 1970s \cite{Penrose:1976}:

\begin{quote}
Despite impressive progress \ldots towards the intended goal of a satisfactory quantum theory of gravity, there remain fundamental problems whose solutions do not appear to be yet in sight. \ldots [I]t has been argued that EinsteinÕs equations should perhaps be replaced by something more compatible with conventional quantum theory. There is also the alternative possibility, which has occasionally been aired, that some of the basic principles of quantum mechanics may need to be called into question. 
\end{quote}
Consistent with this `alternative possibility', it is suggested that quantum theory is not fundamental, but rather should be considered the (singular) limit of a causal deterministic theory of gravity which is geometric not only in space-time but also in state space. That is to say, it is proposed that invariant set theory be considered the basis for a causal `gravitational theory of the quantum'.  As a result, in Section \ref{gravtheory} it is claimed that contemporary programmes, which seek a `quantum theory of gravity' by applying standard field quantisation to some putative gravitational lagrangian, are misguided, erroneously putting the (quantum) cart before the (gravitational) horse. As a specific illustration of the potential of invariant set theory to provide new thinking about contemporary problems in quantum gravity, ideas developed in this paper are applied to the black-hole information paradox, and to the inconsistency between vacuum energy and the cosmological constant. 

In Section \ref{others}, links between invariant set theory and other approaches to quantum physics (decoherence, Bohmian theory, Everettian theory and objective reduction theories) are briefly discussed. In Section \ref{conclusions}, in describing future work, it is concluded that tools from noncommutative geometry may be needed to describe rigorously a version of the Schr\"{o}dinger equation consistent with the invariant set hypothesis. We conclude in Section \ref{conclusions} by returning to Fig \ref{fig:triangle}. 

\section{Some Mathematical Properties of the Lorenz Attractor}
\label{remarkable}

Ed Lorenz's 3-component model of chaos ~\cite{Lorenz:1963}
\begin{eqnarray}
\label{lorenz}
\dot X &=& -\sigma X + \sigma Y \nonumber \\
\dot Y &=& -XZ+rX-Y \nonumber \\
\dot Z &=&  \;\;\;XY -bZ 
\end{eqnarray}
is familiar to most physicists. Lorenz's motivation in deriving this model - a low-order truncation of the partial differential equations for thermal convection in a dissipative fluid - was to show that the concept of long-range weather forecasting using statistical analogue methods was fundamentally flawed. Under the irreversible action of the dynamics (see the discussion in Section \ref{universe} below), state-space volumes converge onto some zero-volume subset of the system's three dimensional state space. But what sort of subset is this? Lorenz agonised about this for some time. Initially he imagined that the attractor of his equations comprised a pair of surfaces which somehow merged at their intersection. However, he rapidly realised this cannot be. In his notes he writes \cite{Palmer:2009b}:
\begin{quote}

 `We see that each surface is really a pair of surfaces, so that, where they appear to merge, there are really four surfaces. Continuing this process for another circuit, we see that there are really eight surfaces etc and we finally conclude that there is an infinite complex of surfaces, each extremely close to one or the other of two merging surfaces.'
 
 \end{quote}
With hindsight, we now understand that Lorenz was describing a fractal geometry $I_L$ in the three dimensional state space spanned by $(X,Y,Z)$. The differential equations (\ref{lorenz}), which the founding father of classical physics, Isaac Newton, would have certainly understood, give rise to a type of nonclassical geometry that would have been utterly alien to Newton. Curiously, one of the founding fathers of fractal geometry was also a meteorologist from an earlier generation: Lewis Fry Richardson \cite{Mandelbrot}.

On the other hand, the notion of chaotic unpredictability attributed to Lorenz and based on the concept of sensitive dependence on initial conditions, predates Lorenz by many decades, going back at least to Poincar\'{e}'s work on the gravitational three-body problem. Indeed Poincar\'{e} himself understood that such sensitive dependence underpinned the unreliability of deterministic weather forecasts. So what was Lorenz's enduring contribution to science? At one level, he gave the world a simple (and remarkably compact) set of ordinary differential equations with which to study chaos. But at a much deeper level, he demonstrated a profound link between the classical space-time calculus of Newton and the nonclassical state-space fractal geometry of Cantor and Richardson. In order to show just how remarkable this link is, we shall relate Lorenzian geometry to two of the most famous theorems of 20th Century mathematics.

Consider a point $p$ in the three-dimensional Lorenz state space. Is there an algorithm for determining whether $p$ belongs to $I_L$? There are certainly large parts of state space which don't contain any part of $I_L$. However, suppose $p$ was a point which `looked' as if it might belong to $I_L$. How would one establish whether this really is the case or not? If we could initialise the Lorenz equations at some point which was known to lie on $I_L$, we could then run (\ref{lorenz}) forward to see if the trajectory passes through $p$. If the integration is terminated after any finite time and the trajectory still hasn't passed through $p$, we can't really deduce anything. We can't be sure that if the integration was continued, it would pass through $p$ at some future stage. 

The Lorenz attractor provides a geometric illustration of the G\"{o}del/Turing incompleteness theorems: not all problems in mathematics are solvable by algorithm. This linkage has been made rigorous by the following theorem \cite{Blum}: so-called Halting Sets must have integral Hausdorff dimension. $I_L$ has fractional Hausdorff dimension - this is why it is called a fractal. Hence we can say that $I_L$ is formally non-computational. To be a bit more concrete, consider one of the classic undecidable problems of computing theory: the Post Correspondence Problem \cite{Sipser}. Dube \cite{Dube:1993} has shown that this problem is equivalent to asking whether a given line intersects the fractal invariant set of an iterated function system \cite{Barnsley}. In general, non-computational problems can all be posed in this fractal geometric way. 

The notion of non-computability is uncommon in physics. However, black-hole event horizons provide a relatively simple and illuminating example of non-computability. Consider the question: Is there a finite algorithm for determining the position in space of a black hole event horizon at some time $t_0$ (i.e. on a spacelike hypersurface labelled by $t_0$)? The answer is no. Like an invariant set, the event horizon is a global concept, but defined in space-time rather than state space (an event horizon is bounding null surface between regions where light rays either escape to infinity or are trapped). As such, the event horizon's position at $t_0$ can be influenced by events arbitrarily in the future of $t_0$ (e.g. whether a massive object falls into the black hole at some $t_1 \gg t_0$). For $t_1$ sufficiently far into the future of $t_0$, no algorithm of given finite length will be able to compute, at $t_0$, whether or not the object will fall into the black hole at $t_1$. Hence the question of determining whether the event horizon passes through a given point in space time is similarly non-computable. 

\begin{figure}
\centering
\includegraphics[scale=0.7]{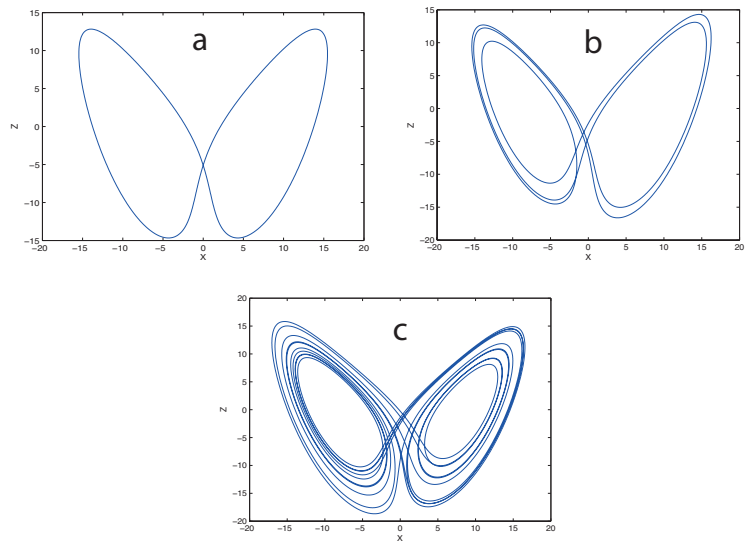}
\caption{Examples of periodic orbits associated with the Lorenz equations (\ref{lorenz}). These can be described by the symbolic bit strings: a) LR, b) LRLRL c) RLLRLLLRRRLLRLRRRRLL}
\label{fig:upo}
\end{figure}

Let us turn to another area of mathematics. Consider, for example, the unstable periodic orbits `embedded' in $I_L$. Fig \ref{fig:upo} shows some examples of such orbits.  If we partition state space so that the region associated with the left-hand wing of the `butterfly' $I_L$ is labelled `L' and the region associated with the right-hand wing `R', then periodic orbits can be described by finite sequences of these labels such as LRLRL, LRLRLRL, or LRLRRRLRRR. In general these periodic orbits are knotted. For example, the orbit `LRLRL' is topologically equivalent to the trefoil knot. On the other hand, there is no periodic orbit of the Lorenz equations which is equivalent to a figure-of-eight knot. Is there some way of characterising the knottedness of all of the unstable periodic orbits of the Lorenz equations (there are infinitely many of them)?

Remarkably, Ghys \cite{Ghys} has shown that the so-called Lorenz knots coincide precisely with the knots of a mathematically much simpler dynamical system called a modular flow.  The modular group $SL(2,\mathbb{Z})/\{I,-I\}$ is the group of $2 \times 2$ matrices with unit determinant (and where $\pm$ the identity element are identified). For example, the matrix

\be
\mathrm A=\left(
\begin{array}{cc}
164&133 \\
127 & 103
\end{array} \right) \nonumber
\ee
is an element of the modular group. Using the `elemental' matrices
\be
\mathrm L=\left(
\begin{array}{cc}
1&1 \\
0 & 1
\end{array} \right) 
\; \; \; \mathrm R=\left(
\begin{array}{cc}
1&0 \\
1 & 1
\end{array} \right) \nonumber
\ee
then, modulo cyclic permutations, A=LRRRLLRRRLLLLR. It turns out that the symbolic sequences of periodic orbits of the Lorenz equations are the same as symbolic sequences of representations of elements of the modular group \cite{Ghys}. The mathematics needed to demonstrate this linkage will be familiar to number theorists - it involves lattices in the complex plane, and associated Eisenstein series and Weierstrass elliptic functions.  In turn, these relates to the mathematics of modular forms and elliptic curves, the mathematics that Andrew Wiles unified in his celebrated proof of Fermat's Last Theorem. This linkage means that if we want to find representations of the periodic orbits of the Lorenz equations to within topological equivalence, we do not need to solve the differential equations (\ref{lorenz}), rather we choose elements of the modular group. Since a trajectory on $I_L$ will eventually return arbitrarily closely to its initial point under the action of the dynamics, symbolic sequences associated with the fractal $I_L$ are composed of these modular strings. 

The `relegation' of differential equations in space-time (the lifeblood of standard theoretical physics) and the corresponding `promotion' of some underlying nonclassical geometry in state space is a primary theme of this paper. The symbolic representation of the invariant set concept, as discussed here, will be central to the mathematical development of these ideas.

\section{The Liouville Equation}
\label{sec:liouville}

In the last section was described some remarkable links between the differential equations (\ref{lorenz}) and two of the deepest theorems of 20th Century mathematics. These links were uncovered by focussing on the state-space geometry $I_L$ associated with these equations, rather than on the differential equations themselves. Is it possible, by focussing on state-space geometry rather than space-time differential equations, we may gain new insights into some of the deepest problems of fundamental physics? By `fundamental physics' we certainly include quantum theory and quantum field theory - the bedrock of standard approaches to the still-sought grand unification of forces in physics. Could these new insights include a reappraisal of the possibility that quantum physics is, after all, underpinned by determinism (no dice) and causality (no spooky action at a distance)? 

The canonical view about the indeterminism of quantum physics has been expressed by Hawking \cite{Hawking}:
\begin{quote}

`According to quantum physics, no matter how much information we obtain or how powerful our computing abilities, the outcomes of physical processes cannot be predicted with certainty because they are not determined with certainty.'

\end{quote}
According to this view there can be no link between quantum theory and determinism, fractal or otherwise, simply because quantum physics is not believed to be deterministic at all. The pioneers of quantum theory were not so unequivocal. Dirac, speaking for himself and his illustrious colleagues,  said \cite{Dirac}:
\begin{quote}

`I must say that [like Einstein and Schr\"{o}dinger] I also do not like indeterminism. I have to accept it because it is certainly the best that we can do with our present knowledge. One can always hope that there will be future developments which will lead to a drastically different theory from the present quantum mechanical theory and for which there may be a partial return to determinism.'
\end{quote}
Could determinism on non-computable fractals provide the `partial' return of determinism for which Dirac hoped? Some support for this is provided by Penrose \cite{Penrose:1989} who says:
\begin{quote}

`It seems to me to be quite plausible that the correct theory of quantum gravity might be a deterministic but non-computable theory.'

\end{quote}

At the level of differential equations such as (\ref{lorenz}), the notion that deterministic chaos could explain the unpredictability of quantum measurement would appear to be nonsensical: chaotic equations are nonlinear whilst the Schr\"{o}dinger equation (in either non-relativistic or relativistic form) is linear. How could a focus on state-space geometry help when trying to bridge this seemingly unbridgeable gap?

\begin{figure}
\centering
\includegraphics[scale=0.4]{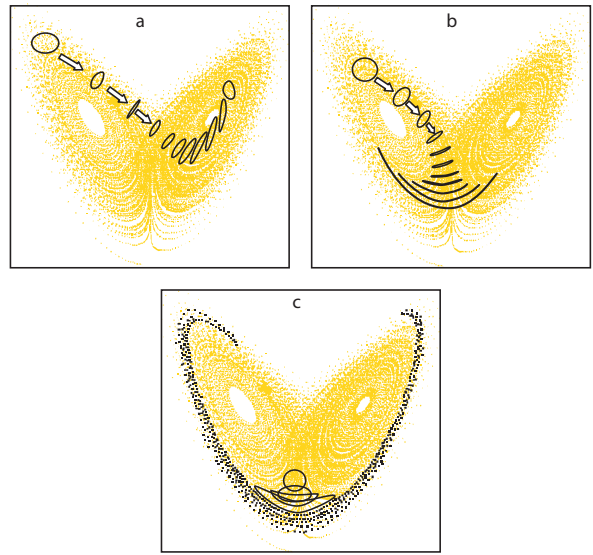}
\caption{The top left figure illustrates evolution of uncertainty on the Lorenz Attractor $I_L$ for a) a relatively stable part of $I_L$, b) a more `typical' situation, c)  a particularly unstable part of $I_L$.}
\label{fig:lorenz}
\end{figure}

Fig \ref{fig:lorenz} illustrates three instances of the evolution close to $I_L$ of a set of points arranged in a `ring' around some central point. The ring characterises some uncertainty in the initial state. A number of points can be made. Firstly, the three panels together illustrate the nonlinearity of the underlying equations (\ref{lorenz}); if we write (\ref{lorenz}) in the general form $\dot X= F[X]$ and linearise about a particular state $X$, then the equation which describes the growth of small perturbations $\delta X$ about $X$ can be written $\delta \dot X = dF/dX \delta X$ where  $dF/dX$ is the so-called Jacobian operator. Now because $F$ is nonlinear, it is at least quadratic in $X$, which means that the Jacobian cannot be independent of $X$. We see in Fig \ref{fig:lorenz} the dependence of the growth of small perturbations on X. 

The initial ring of points in each panel in Fig \ref{fig:lorenz} samples a contour of equal values of a distribution or measure $\rho$, giving the probability that the `true' initial state lies within that contour.  The equation which describes the evolution of $\rho$ is the classical Liouville equation 
\be
\label{liouville}
\frac{\partial \rho}{\partial t}+\nabla . (\rho \mathbf{v})=0
\ee
Here $\mathbf{v}$ denotes a state-space velocity vector, equal to $(\dot X, \dot Y, \dot Z)$ for the Lorenz system. The Liouville equation describes conservation of probability. Hence if there is a 90\% probability that the true initial state lies within a particular contour at initial time, there is a 90\% probability that the true state at some future time $t$ lies within that same contour evolved to $t$. Note that even though the deterministic equations which underpin the evolution of states e.g. (\ref{lorenz}) are nonlinear, the Liouville equation is itself precisely linear in $\rho$. Unfortunately, this doesnÕt make this (effectively infinite dimensional) equation easy to solve, even for relatively low-order dynamical systems \cite{Ehrendorfer:2006}. In the case where state evolution is Hamiltonian (or can be well approximated as Hamiltonian), the Liouville equation (\ref{liouville}) can be written in the form
\be
\label{liouvillehamilton}
\frac{\partial \rho}{\partial t}+[\rho,H]=0
\ee
where the bracket $[\dots, \ldots]$ denotes the well-known Poisson Bracket \cite{Goldstein}. In should be noted in passing that in Hamiltonian theory, what here is called `state space', is often referred to as `phase space'. For the present purposes, the two terms can be considered synonymous.  

Suppose it was known that, although uncertain, the state certainly lies on the invariant set $I_L$. The evolution of uncertainty will still be governed by a Liouville equation. However, the geometry of $I_U$ is too irregular to allow us to use the simple calculus as in the second term in (\ref{liouville}). One would need a mathematical structure which is general enough to allow the irregular non-differentiable geometric properties of $I_L$. Even though fractal geometries such as $I_L$ are commutative, the mathematics of non-commutative geometry is general enough to take on this role \cite{Connes} \cite{Lapidus}. A discussion of the mathematics of non-commutative geometry is beyond the scope of this paper. 

We can compare the classical Liouville equation with the Schr\"{o}dinger equation written in the form
\be
\label{schrodinger}
i\hbar\frac{\partial \rho}{\partial t}+[\rho,H]=0
\ee
where now $\rho$ is the density matrix operator and $[\ldots, \ldots]$ denotes the operator commutator in complex Hilbert Space. The structure of the Schr\"{o}dinger equation in the form (\ref{schrodinger}) is remarkably similar to that of the classical Liouville equation (\ref{liouvillehamilton}). Hence, just as the linearity of the classical Liouville equation says nothing about the nonlinearity of the underpinning deterministic equations from which the probability distribution $\rho$ is derived, could it be that the linearity of the Schr\"{o}dinger equation may not in any way imply the non-existence of a deeper deterministic description of physics which is profoundly nonlinear?

Of course there are differences between the Liouville equation and the Schr\"{o}dinger equation \cite{CavesSchack}.  For example, a square root of minus one times Planck's constant multiplies the partial time derivative in the latter. Also, the dependent variables of the Schr\"{o}dinger equation are not simple functions in some Euclidean state space, but are operators in a complex Hilbert Space. Are these differences important? Yes, for sure! Because of these differences there are a number of `no-go' theorems which seemingly prevent (\ref{schrodinger}) from being interpreted as a Liouville equation. The most famous of these no-go theorems is the Bell Theorem, discussed in Sections \ref{belltheorem} and \ref{emergent} below (which asserts that models based on some realistic definition of state must necessarily violate local causality). Another is the Kocken-Specker \cite{KochenSpecker} theorem also discussed briefly in Section \ref{emergent} (which asserts the impossibility of assigning values to all physical quantities whilst, at the same time preserving the functional relations between them). 

Are these `no-go' theorems showstoppers, as far as the (deterministic/causal) aspirations of Dirac, Penrose and others are concerned? Overwhelmingly the view of the physics community is that they are. In the coming sections, it is discussed why, by treating fractal (measure-zero) state-space geometry as a primitive concept in fundamental physics, these `no-go' theorems may not be showstoppers at all!

\section{The Universe as a Dynamical System}
\label{universe}

The first step in our search for a causal deterministic description of quantum physics is to link some of the dynamical systems discussion above with modern ideas from the field of cosmology and general relativity. First, consider the universe $U$ as a nonlinear dynamical system \cite{Wainwright}. Let $p$ denote a point in the state space of $U$. Then the values of some set of coordinates which span that space will provide all the information needed to determine the state of the universe at some moment in time. A trajectory in state space therefore describes a space-time: both the geometry of the space-time and the evolution of all the matter that exists in that space-time. Following recent developments (e.g. \cite{SteinhardtTurok} and \cite{Penrose:2010}), we consider quasi-cyclic cosmologies evolving through a series of aeons from one big bang to the next. 

Given such quasi-periodicity, is it possible that $U$ lies on a zero-volume fractal subset $I_U$ of some compact volume in state space (similar to $I_L$)? A key point here is that in nonlinear dynamical systems' theory, such zero volume sets arise when the dynamics are irreversible (i.e. dissipative). To see this, consider the Lorenz equations (\ref{lorenz}). Under their action, some comoving volume $V$ in state space evolves according to 
\be
\frac{dV}{dt} = \int_{\partial V} \mathbf v . dS \nonumber
\ee
where, again, $\mathbf v =(\dot X, \dot Y, \dot Z)$ and $\partial V$ is the boundary of $V$. From Gauss' theorem and the fact that the system parameters $\sigma$ and $b$ are each positive numbers,
\be
\label{convergence}
\frac{dV}{dt} = \int_V \nabla \cdot \mathbf v \;dv =-(\sigma +b+1) V < 0
\ee
Hence $V(t) \rightarrow 0$ as $t \rightarrow \infty$. An important point (for all that follows) is that such zero volume fractal invariant sets cannot arise in Hamiltonian chaos. For such systems, Liouville's theorem states that $\nabla \cdot \mathbf v= 0$, implying no shrinkage of volume. (It is important to note that the Liouville equation (\ref{liouville}) describes conservation of probability even when $\nabla \cdot \mathbf v \ne 0$.)

If $U$ lies on a zero volume invariant set $I_U$, there must be some cosmological equivalent of the state-space convergence $\nabla \cdot \mathbf v<0$. What could this be? To begin to answer this, consider the problem of the  `Hawking Box' discussed extensively by Penrose \cite{Penrose:1989} \cite{Penrose:1994} \cite{Penrose:2004} \cite {Penrose:2010}. The problem is to describe the asymptotic evolution of an isolated distribution of matter sufficiently large that it can collapse to one or more black holes. When formed, such black holes will re-radiate this matter back into space through Hawking radiation. Because of the link to black holes, it is convenient to partition the state space of the Hawking Box into a region $\mathcal{B}$ containing black holes, and a complement $\mathcal{B}'$ free of black holes. A key element of Penrose's discussion of the `phase flow' of the Hawking Box concerns the role of black hole information loss, which can be considered a consequence of the black-hole no-hair theorem. As discussed in Section \ref{gravtheory} below, this notion of information loss is a controversial one \cite{Susskind:2008} (and new ideas will be presented below to try to resolve the controversy). However, to quote Penrose \cite{Penrose:2010}
\begin{quote}
`A better way of describing this [information loss] is as a loss of degrees of freedom, so that \ldots the phase space [of the universe] has actually become smaller than it was before. '
\end{quote}
Penrose argues that this process of state-space convergence is critical if we are to account for the second law of thermodynamics in an oscillatory universe. 

By analogy with (\ref{convergence}), consider the asymptotic evolution of some hypothetical volume $V$ in the state space of $U$. We will assume the process described by Penrose i.e. that $\nabla \cdot \mathbf v< 0$ in the region $\mathcal{B}$ of the state space of $U$. Hence, $V$ will eventually shrink to nothing. This does not immediately imply that the asymptotic state of $V$ is a fractal. Measure-zero fractal invariant sets are generic features of forced dissipative dynamical systems. In multi-scale dynamical systems (e.g. describing turbulent Navier-Stokes flow) the dissipation occurs on small scales whilst the forcing occurs on large scales. The forcing and dissipation must be in some overall balance in order that the invariant set is structurally stable.  If the asymptotic state of $V$ is a fractal, and if Planck-scale black-hole information loss provides the process which mimics small-scale dissipation, what is the process that might mimic the corresponding large-scale forcing? As discussed further in Section \ref{gravtheory}, it can be speculated that the recently discovered positive cosmological constant, which is accelerating the universe, is this large-scale forcing. As such, the cosmological constant can be neither too small, nor too large. Too small and the invariant set becomes a fixed point or completely periodic limit cycle (with no associated probability structure). Too large and the system and has no (compact) invariant set structure at all. That is to say, the balance described below, where the invariant set has enough fractal structure to allow a description of quantum probability, but not so much that it is constrained by the Bell inequality, may require some subtle balance between large and small scale processes acting in the universe. 

By analogy with the theory of forced dissipative dynamical systems theory (but, to repeat, not Hamiltonian theory) it is plausible that $V$, left to evolve over countless aeons, will evolve to a zero-volume fractal invariant set $I_U$ in the state space of the universe. However, the notion of having to wait an infinite number of aeons for this to happen is awkward from a physical point of view. Hence one can simply postulate as a (new) primitive law of physics that the state of the universe lies \emph{precisely} on the fractal invariant set $I_U$ (and hence has lain on $I_U$ for all past time and will lie on $I_U$ for all future time). This is referred to as the `Cosmological Invariant Set Postulate' \cite{Palmer:2009b}. Like the event horizon of a black hole, the invariant set concept is causal, atemporal and non-computable. 

The notion of the universe evolving on a fractal invariant set implies a rather novel perspective on the multiverse, a concept so prevalent in modern cosmology. That is to say,  the `parallel universes' in the neighbourhood of some $p \in I_U$ do not actually represent `other worlds' at all, but rather represent states of our own world at future aeons (when the state of $U$ has returned to points close to $p$). The state-space closeness of these neighbouring states, given their remoteness in terms of conventional `time', could be seen as a manifestation of the distinction Bohm made between implicate and explicate order \cite{Bohm}. \footnote{It is interesting to note that Bohm's ideas about implicate and explicate order were inspired by watching the stretching and folding of treacle, similar to the essential stretching and folding of the Smale horseshoe map, a canonical mapping for generating chaotic invariant sets in state space}. 

\section{Nonlocality and the Bell Theorem}
\label{belltheorem}

We are now in a position to begin a discussion of what is generally considered to be the principal reason why quantum physics cannot be described both realistically and causally. Bell's Theorem \cite{Bell} is usually interpreted as saying that no physical theory based on locally causal hidden variables can ever reproduce all the predictions of quantum mechanics. 

Using Bohm's adaptation of the famous EPR experiment (e.g. \cite{Shimony}), consider a pair of spin-1/2 particles in the superposed singlet spin state (using quantum mechanical language). Consistent with the invariant set postulate, we assume local realism. Hence, if particle 1 is measured  in the $\mathbf{\hat a}$ direction, the outcome is given by $A(\mathbf{\hat a}, \lambda)=\pm 1$, with an analogous function $B(\mathbf{\hat b}, \lambda)=\pm 1$ for particle 2, where the auxiliary (or hidden) variable $\lambda$ labels the elements of reality of the composite system 1+2. By definition, the result $B$ for particle 2 does not depend on the setting $\mathbf{\hat a}$ of the magnet for particle 1, nor $A$ on $\mathbf{\hat b}$. According to Bell's definition \cite{Bell}, a theory can be said to be locally causal if the value of an observable (what Bell calls a `beable') at some point  $p$ in space-time is determined by the values of observables within the past light cone of $p$. The assumption $A=A(\mathbf{\hat a}, \lambda), B=B(\mathbf{\hat b}, \lambda)$ is consistent with local causality. Conversely, if it were the case that $B$ depended on the setting $\mathbf{\hat a}$, and the devices measuring the two particles were sufficiently remote, then local causality would certainly be violated. As endless commentators have noted, any violation of local causality would sit uneasily with relativity theory. 

The correlation of outcomes of spin measurements in the $\mathbf{\hat a}$ and $\mathbf{\hat b}$ directions on a sample of such particles 1 and 2 is given by
\be
\label{correlation}
Corr_{\rho}(\mathbf{\hat a}, \mathbf{\hat b})=\int_{\Lambda} A(\mathbf{\hat a}, \lambda) B(\mathbf{\hat b}, \lambda) d\rho
\ee
where $\Lambda$ is some suitably complete probability distribution over $\lambda$ (about which more will be written below). Now for all $\lambda$ and $\mathbf{\hat c}$, 
\be
\label{opposite}
A(\mathbf{\hat c}, \lambda)= - B(\mathbf{\hat c}, \lambda).
\ee
From (\ref{opposite}) and (\ref{correlation}) Bell derives the eponymous inequality
\be
\label{bell}
|Corr_{\rho}(\mathbf{\hat a}, \mathbf{\hat b})-Corr_{\rho}(\mathbf{\hat a}, \mathbf{\hat c})| \le 1+Corr_{\rho}(\mathbf{\hat b}, \mathbf{\hat c})
\ee
Correlations which violate this (or the alternative Clauser Horn Shimony and Holt -CHSH\cite{CHSH}) inequality are generally referred to as `nonlocal correlations' \cite{Bancal:2013}, but (so as not to prejudge the matter) we refer to them here as `nonclassical correlations' instead. There is no doubt whatsoever that, consistent with quantum theory, experimentally determined correlations are nonclassical in the sense described above \cite{Aspect:1981}. These experiments are completely robust and not sensitive to particular experimental choices of parameters $\mathbf{\hat a}$ and $\mathbf{\hat b}$. 

Certainly $I_U$ is both realistic and causal. Does this mean it is necessarily constrained by the Bell inequality? No. For a putative deterministic causal theory to be constrained by the Bell inequality (\ref{bell}), it has additionally to be assumed that for any particular $\lambda \in \Lambda$, all three of the triples $(\mathbf{\hat a}, \mathbf{\hat b}, \lambda)$, $(\mathbf{\hat a}, \mathbf{\hat c}, \lambda)$ and $(\mathbf{\hat b}, \mathbf{\hat c}, \lambda)$ exist, i.e. all three triples correspond to states of physical reality. Without this assumption, the inequality (\ref{bell}) could manifestly not be derived. This assumption can be considered a particular case of the so-called measurement independence postulate (eg \cite{Hall:2010}) -  that the probability $\rho(\lambda|\mathbf{\hat a}, \mathbf{\hat b})=\rho(\lambda|\mathbf{\hat a'}, \mathbf{\hat b'})$ for all $\mathbf{\hat a}, \mathbf{\hat b},\mathbf{\hat a'}, \mathbf{\hat b'}$.

A universe where the measurement independence postulate is maximally violated (ie where a particular $\lambda$ is uniquely associated with a specific pair of measurement orientations $\mathbf{\hat a}, \mathbf{\hat b}$) is usually referred to as `superdeterministic'. The state space of a superdeterministic universe is a lone trajectory with no neighbours (in the present dynamical system's context, this could be associated with an invariant set which was a simple limit cycle). In particular, this state space has no natural structure with which to define alternative counterfactual worlds where, for a given $\lambda$, the experimenters might have chosen alternative measurement orientations to the ones they actually chose. For this reason, the concept of probability has to be `put in by hand' into a superdeterministic theory, and inevitably this appears extremely artificial and hence implausible (Bell \cite{Bell} used the word `conspiratorial' to describe the emergence of quantum probability in a superdeterministic world). As a result most physicists, quite rightly in the authors' view, find this type of extreme superdeterminism to be completely unacceptable as a theory of fundamental physics. 

However, as has recently been shown \cite{Hall:2010} \cite{Hall:2011}, it is not necessary to maximally violate the measurement independence postulate in order to explain quantum entanglement correlations; explaining quantum singlet correlations in a causal deterministic framework requires just $1/15$ of one bit of correlation between the measurement settings and $\lambda$. This raises the question as to whether there exists a cosmological invariant set whose state-space structure is on the one hand rich enough to allow (quantum) probabilities to be defined naturally (ie without conspiracy and without restricting experimenter choice in any practical sense), but is on the other hand sparse enough to allow a partial violation of the measurement independence postulate, sufficient to account for the experimental violation of the Bell inequality. It is argued in Section \ref{thebelltheorem} that there does exist such a cosmological invariant set.  

\section{Quantum Physics and Invariant Set Theory}
\label{quantum}

In the last section, we discussed the possibility of nullifying the Bell Theorem using the cosmological invariant set concept. In this Section we construct a fractal invariant set (a prototype for $I_U$), using the type of symbolic representations discussed in Section \ref{remarkable}, for which this nullification can become an actuality. In constructing this symbolic representation, we focus on the three key differences highlighted in Section \ref{sec:liouville} between the classical Liouville equation and the Schr\"{o}dinger equation:
\begin{itemize}
\item $i$
\item $\hbar$
\item The complex Hilbert Space
\end{itemize}
and show how these arise naturally from the cosmological invariant set postulate. 

\subsection{Cantor sets and square roots of minus one}
\label{complex}

\begin{figure}
\centering
\includegraphics[scale=0.7]{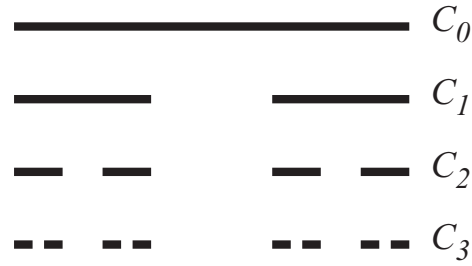}
\caption{The first four approximations of the famous `middle-thirds' Cantor Set $C$.}
\label{fig:cantor}
\end{figure}

Locally, invariant sets such as $I_L$ can be written as $C \times \mathbb{R}$ where $C$ is a multidimensional Cantor Set and $\mathbb{R}$ denotes a trajectory segment in state space.  Let us therefore start by considering the simplest fractal set, the Cantor ternary set $C$, which can be defined as
\be
C=\bigcap_{k \in \mathbb{N}} C_k \nonumber
\ee
where the $C_k$ denote iterative approximations to $C$ (see Fig \ref{fig:cantor}). One can represent a point in $C$ as a real number $0 \le r \le 1$ whose base-3 representation contains no digit 1. Hence $.02002020222$ is an element of $C$ but $.0221022002$ is not. Associated with each individual interval of $C_{k-1}$, the intervals of $C_{k}$ can therefore be labelled by two integers e.g. $0$ and $2$. 

In order to construct fractal sets from which the properties of quantum qubit physics are emergent, it is necessary to generalise this particular `middle-thirds' rule. In particular, consider a family of Cantor Sets
\be
\label{CN}
C^{(N)}=\bigcap_{k \in \mathbb{N}} C^{(N)}_k \nonumber
\ee
where $C^{(N)}_k$ comprises $2^N(2^N+1)$ copies of $C^{(N)}_{k-1}$ (see Appendix A for more details). Fig \ref{fig:cantor2} shows two different possible representations of $C^{(2)}_k$ associated with a given interval of $C^{(2)}_{k-1}$. These two representations, labelled by $t_i$ (Fig \ref{fig:cantor2} middle line) or $t_f$  (Fig \ref{fig:cantor2} bottom line) have the same fractal dimension, but differ in terms of what Mandelbrot \cite{Mandelbrot} has referred to as `lacunarity'. Hence in Fig \ref{fig:cantor2}, $C^{(2)}(t_i)$ comprises 5 uniformly spaced groupings of 4 intervals with relatively small lacunarity, whilst $C^{(2)}(t_f)$ comprises 2 groupings of 10 intervals, having relatively large lacunarity. The two groupings of intervals of $C^{(2)}_k(t_f)$ are labelled as `$a$' (and coloured red) and `$\lnot a$' (and coloured black) respectively. 

A generic binary labelling of any one of the 5 groupings of intervals of $C^{(2)}_k(t_i)$ can be written as
\be
a_1 a_2 a_3 a_4 \nonumber
\ee
where $a_j \in \{a, \lnot a\}$. We now use a permutation/negation representation of complex numbers to define a deterministic rule whereby each interval of $C^{(2)}_k(t_i)$ is also labelled `$a$' (red) or `$\lnot a$' (black): in Fig \ref{fig:cantor2} red arrows link red intervals in $C^{(2)}_k(t_i)$ with red intervals of $C^{(2)}_k(t_f)$ and similarly for black arrows. Define the permutation/negation operator $i$ by
\be
\label{i}
i(a_1 a_2 a_3 a_4)=a_2 \lnot a_1 a_4 \lnot a_3 
\ee
where $\lnot(\lnot a)=a$. Applying (\ref{i}) twice, it is very easily shown that
\be
\label{sq}
i \circ i(a_1 a_2 a_3 a_4)=i^2(a_1 a_2 a_3 a_4)=\lnot a_1 \lnot a_2 \lnot a_3 \lnot a_4\equiv -(a_1 a_2 a_3 a_4)
\ee
so that $i$ can be considered a `square root of minus one'. Moreover, with 
\be
\label{sqsq}
i^{1/2}(a_1 a_2 a_3 a_4)=a_3 a_4 a_2 \lnot a_1
\ee
then $i^{1/2}$ is a `square root of $i$' (so $i^{1/2} \circ i^{1/2} = i$). Let $P_a[i^\alpha(aaaa)]$ denote the frequency of occurrence of the symbol `$a$' in the grouping $i^{\alpha}(aaaa)$, where $0 \le \alpha \le 4$ is an integer multiple of $1/2$. Then 
\be
\label{prob}
P_a[i^\alpha(aaaa)]=|1-\frac{\alpha}{2}|
\ee
$P_a$ is referred to as a symbolic probability function. All 5 groupings of $C^{(2)}_k(t_i)$ in Fig \ref{fig:cantor2} can be labelled as
\be
\label{grouping}
i^0(aaaa)\;\;\;\;i^{1/2}(aaaa)\;\;\;\; i(aaaa)\;\;\;\; i^{3/2}(aaaa)\;\;\;\;i^2(aaaa)
\ee
where $i^0 \equiv \text{id}$ and $i^{3/2}=i \circ i^{1/2}$. Notice, consistent with Liouville evolution, the mapping given by the red and black arrows in Fig \ref{fig:cantor2} preserves probability. For example, since equal numbers of intervals of $C^{(2)}_k(t_i)$ have `$a$' and `$\lnot a$' labels, so too do intervals of $C^{(2)}_k(t_f)$. 

The construction above is readily generalised for arbitrary $N>0$ \cite{Palmer:2012}. For all $N$ $C^{(N)}_k(t_f)$ continues to have just two groupings (labelled $a$ and $\lnot a$ as before). However, the larger is $N$ the bigger the set of fractional roots of $i$ used to label the multiple groupings of $C^{(N)}_k(t_i)$.  Hence for large $N$, the mapping $\mathcal D: C^{(N)}(t_i) \rightarrow C^{(N)}(t_f)$, as defined by these roots of $i$, is a bijection between a quasi-uniform fractal and a highly lacunar fractal. As will be discussed in more detail in Section \ref{planck}, the two groupings of  $C^{(N)}_k(t_f)$ are to be considered gravitational attractors on $I_U$ and define discrete measurement outcomes in qubit quantum physics. The labelling of $C^{(N)}_k(t_i)$ defines the (generically riddled \cite{Alexander:1992}) basins of attraction into which the intervals of $C^{(N)}_k(t_i)$ belong.  

\begin{figure}
\centering
\includegraphics[scale=0.8]{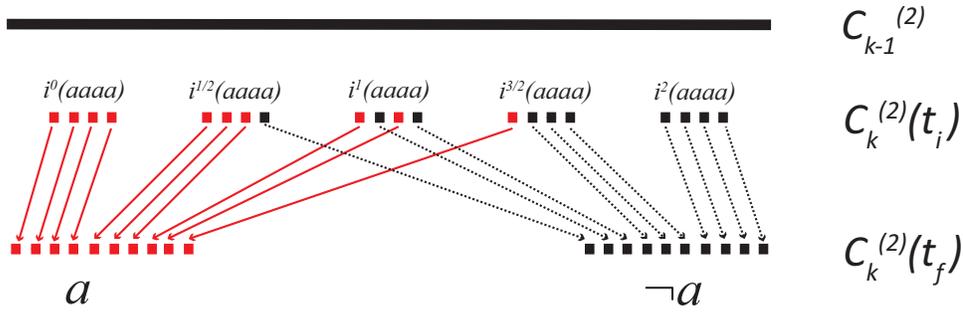}
\caption{For a given interval of the $k-1$th approximation for the Cantor Set $C^{(2)}$ (top row), two different $k$th approximations are illustrated with low and high fractal lacunarity (middle and bottom row respectively). The intervals of $C^{(2)}_k(t_f)$ are labelled `$a$' or `$\lnot a$' according to which of the two groupings an interval belongs. Each interval of $C^{(2)}_k(t_i)$ is labelled `$a$' (and coloured red) or `$\lnot a$' (and coloured black) according to a permutation/negation representation of the complex numbers. Symbolic representation of the riddled-basin dynamical evolution from $C^{(2)}_k(t_i)$ to $C^{(2)}_k(t_f)$ is indicated by arrow-headed lines. A generalisation from $C^{(2)}$ to $C^{(N)}$, $N \gg 0$, provides a representation of both complex numbers and quantum measurement in invariant set theory. As discussed in the text, it is proposed that this evolution is fundamentally gravitational in origin.}
\label{fig:cantor2}
\end{figure}

The operator $i$ defined by (\ref{i}) above is one of many square roots of minus one. For example, the operators defined by
\begin{eqnarray}
\label{quaternion}
\mathbf{E_0}(a_1 a_2 a_3 a_4)=a_4 \lnot a_3 a_2 \lnot a_1 \nonumber \\
\mathbf{E_1}(a_1 a_2 a_3 a_4)=a_2 \lnot a_1 \lnot a_4 a_3 \nonumber \\
\mathbf{E_2}(a_1 a_2 a_3 a_4)=a_3 a_4 \lnot a_1 \lnot a_2
\end{eqnarray}
not only satisfy $\mathbf{E}^2_j=-1$), but collectively satisfy the familiar rules for quaternionic multiplication e.g.
\be
\label{quat}
\mathbf{E}_0 \circ \mathbf{E}_1 = \mathbf{E}_2
\ee
It is left as an exercise for the reader to show that the Pauli spin matrices can be written in terms of $\mathbf{E}_j$ and the square-root-of-minus-one representation $i$, expressed as matrices. 

As outlined in Appendix B (see also \cite{Palmer:2012}), this construction is readily generalised to construct sets $\{\mathbf{E}_\beta\}$ of independent quaternion operators acting on symbolic strings of length $2^N$, where $\beta$ is drawn from the set of dyadic rational numbers (between 0 and 4) describable with $N$ bits. Again, as outlined in Appendix B, this can be generalised further to include fractional powers $\mathbf{E}^\alpha_\beta$, where again $\alpha$ is a dyadic rational numbers between 0 and 4 and describable with $N$ bits. Consistent with (\ref{prob})
\be
\label{pf}
P_a[\mathbf{E}^\alpha_\beta(aaa\ldots a)]=|1-\frac{\alpha}{2}|
\ee
The fractional quaternionic permutation/negation operators $\mathbf{E}^\alpha_\beta$ are needed to label multi-dimensional generalisations of $C^{(N)}_k(t_i)$, required to describe multi-qubit physics. Using a metric on the bit strings (eg the Hamming distance), a crucial remark for all that follows is that as $N \rightarrow \infty$, the bit strings $\mathbf{E}_\beta (aaa \ldots a)$ do not vary continuously with $\beta$. Hereafter, the superscript `$(N)$' is dropped, but with the assumption that $N \gg 0$. 

\begin{figure}
\centering
\includegraphics[scale=0.4] {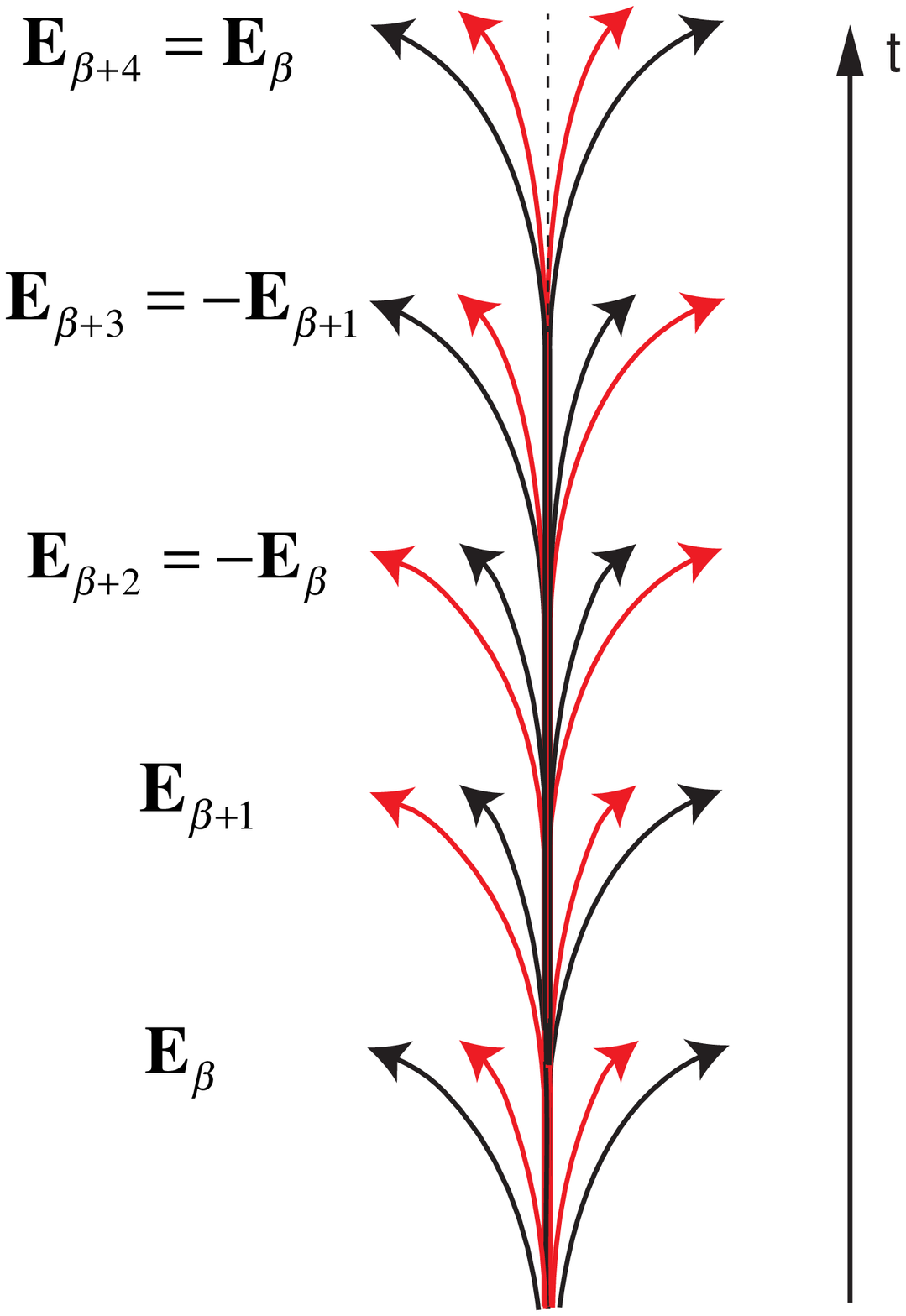}
\caption{The exponential state-space divergence of neighbouring trajectories in time from some fiducial trajectory (dashed line), and the quaternionic  labelling of successive levels of approximation of the invariant set (as defined in Section (\ref{complex})), leads naturally to oscillations in time. By analogy with the geodesic equation in general relativity, invariant set theory's interpretation of  $E=\hbar \omega$ suggests that the quantity we call the `energy' of a particle on the fiducial central trajectory is defined by the divergence of neighbouring trajectories, as measured by the invariant set's Lyapunov exponents.}
\label{fig:oscillate}
\end{figure}

We conclude this section with a brief discussion of one of the key consequences of the role of complex numbers in quantum theory, $E=\hbar \omega$. Consider some fiducial trajectory on $I_U$, e.g. as shown in Fig \ref{fig:oscillate}. Since the dynamics on $I_U$ are assumed chaotic, neighbouring trajectories will in general diverge exponentially from the fiducial trajectory. It is possible to use the permutation/negation operators $\mathbf{E}_\beta$ associated with different levels of iteration $k$, to reveal a temporal oscillatory structure in the labelling patterns, as higher-and-higher approximations $C_k$ of the Cantor Set $C$ are exponentially amplified and `brought to the fore'.  

At this point it is worth considering a correspondence with general relativity, where space-time geometry is considered a primitive concept. In general relativity, the geodesic equation
\be
\ddot x^c + \Gamma^c_{ab} \dot x_a \dot x_b=0
\ee
describes how inertial properties of a test particle moving on some fiducial space-time geodesic are defined by the geometry of space-time in the neighbourhood of this trajectory. By the Jacobi equation for geodesic deviation, this neighbourhood geometry can in turn be estimated from the deviation of geodesics neighbouring the fiducial geodesic That is to say, in general relativity, the kinetic energy of a test particle on the fiducial geodesic can be determined by the divergence of neighbouring geodesics from the fiducial geodesic. 

As discussed, the cosmological invariant set postulate attempts to extend Einstein's insights about the primacy of geometry from space-time to state space. Hence, by extending the argument above into state space, let us assume that the energy $E$ of a particle associated with a fiducial trajectory on $I_U$ is determined by the state-space divergence of neighbouring trajectories from the fiducial trajectory on $I_U$. By virtue of the log-oscillatory structure illustrated in Fig \ref{fig:oscillate} we can therefore consider writing $E=\hbar \omega$ where $\omega$ is proportional to the relevant Lyapunov exponent associated with the rate of neighbouring trajectory divergence. 

If one relates the labelling in Fig \ref{fig:cantor2}, which defines an oscillatory structure in configuration space and the labelling in Fig \ref{fig:oscillate}, which defines an oscillatory structure in the neighbourhood of the fiducial trajectory in time, then Figs \ref{fig:cantor2} and \ref{fig:oscillate} can be seen as two sides of the same coin, as would be required by Lorenz invariance. It can be remarked that, using the self-similar construction in Appendix B, $\mathbf{E}_\beta$ can be easily related to Dirac matrices, suggesting more direct links to Dirac's relativistic form of the Schr\"{o}dinger equation than to the non relativistic Schr\"{o}dinger equation. However, for lack of space, these ideas will have to be pursued elsewhere. 

\subsection{The Complex Hilbert Space}
\label{hilbert}

What is the relationship between the space of symbolic sequences $\mathbf{E}^\alpha_\beta(aaa\ldots a)$ and the complex Hilbert Space of a quantum qubit: the Bloch Sphere? The elements of a Hilbert Space are normalised vectors and it should be recalled that a unit vector is a natural way to represent probability assignments even in elementary classical physics. Consider a two dimensional space, spanned by the $x$-$y$ axes. Let $\mathbf i$, $\mathbf j$ be unit vectors along the $x$ and $y$ directions respectively. If $P_a$ denotes the (frequentist-defined) probability of event $a$ so that $1-P_a$ is the probability of the mutually exclusive event $\lnot a$, then by Pythagoras's theorem the vector
\be
\label{samplespace}
\mathbf{v}(P_a)=\sqrt{P_a}\; \mathbf{i}+\sqrt{1-P_a} \;\mathbf{j}
\ee
has unit norm, and can therefore represent a sample space associated with any probability assignment $P_a$. With (\ref{samplespace}) in mind, we can therefore make the assignments
\be
\label{correspondence}
\mathbf{E}^\alpha_\beta(aaa\ldots a) \sim \cos\frac{\theta}{2}\;|a\rangle + \sin\frac{\theta}{2} e^{i \phi}\;|\lnot a\rangle
\ee
between sample spaces of symbolic symbols and unit vectors in complex space, where
\begin{align}
\label{incom}
\cos^2 \theta/2&= |1-\alpha/2| \nonumber \\
\phi&=\pi\beta/2 
\end{align}
implying that $\cos\theta$ and $\phi/\pi$ must be describable by $N$ bits and hence be dyadic rational. From (\ref{pf}) and (\ref{incom}), the probability that a given element of the symbol sequence $\mathbf{E}^\alpha_\beta(aaa\ldots a) $ is the label $a$, is equal to $\cos^2 \theta/2$. 

This association between symbolic sequences and vectors is straightforward. So why is the Hilbert Space of quantum theory so conceptually problematic? The problem is that Hilbert Space is a complete vector space - indeed the consequent continuity of Hilbert Space can be made an essential axiom of quantum theory \cite{Hardy:2004}. Hence Hilbert Space vectors are defined not only when the squared amplitudes of the basis vectors are rational (as would be the case in (\ref{samplespace}) and (\ref{correspondence})), but also when they are irrational. In invariant set theory, by contrast, these vectors are undefined when the squared amplitudes are irrational. 

Now although it is straightforward to use well-established 19th Century mathematics to complete the dyadic rationals $\cos\theta$ and $\phi/\pi$ into the space of real numbers, we cannot infer from (\ref{correspondence}) that the corresponding complex Hilbert Space of quantum theory is the abstract completion of the space of symbolic sequences. As mentioned above, the bit strings $\mathbf{E}_\beta(aaa\ldots a)$ do not depend continuously on $\beta$. Consider, for example, a Cauchy sequence $\{\beta_k\}$ of dyadic rationals $\beta_k$ whose limit is $\sqrt2$. Associated with each $\beta_k$ is a symbolic sequence $\mathbf{E}_{\beta_k}(aa \ldots a)$ and, by (\ref{correspondence}), an assignment to an element of a Hilbert Space. However, because the sequence $\{\mathbf{E}_{\beta_k}(aa \ldots a)\}$ is not convergent \cite{Palmer:2012}, we cannot assign $ |a\rangle + e^{i \pi/\sqrt 2}\;|\lnot a\rangle$ to the limit of $\{\mathbf{E}_{\beta_k}(aa \ldots a)\}$ in the conventional mathematical sense of the word `limit'. Rather, $|a\rangle + e^{i \pi/\sqrt 2}\;|\lnot a\rangle$ would have to be described as a singular limit \cite{Berry}.

Usually a singular limit denotes a change in the type of an equation, labelled by a parameter, as some parameter limit is reached. For example in the inviscid limit of fluid mechanics (where the inverse Reynolds number goes to zero), the dynamical equations change from parabolic to hyperbolic type. In our discussion of quantum physics, we note a similar change of type: in the irrational limit of the phase angle $\phi/\pi$, the mathematical object describing the notion of `state' is an element of an abstract vector space which cannot be linked to any underpinning frequentist (bit string) sample space. No wonder quantum theory, whose state space is the full Hilbert Space, can appear paradoxical from a physical point of view! Michael Berry \cite{Berry} has argued that singular limits abound in physics and are fundamental in the description of nature at different levels. In particular Berry concludes:
\begin{quote}

 `... there are both reassuring and creative aspects to singular limits. And by regarding them as a general feature of physical science, we get insight into two related philosophical problems: how a more general theory can reduce to a less general theory and how higher-level phenomena can emerge from lower-level ones.' 
\end{quote}
Of course, just as inviscid fluid dynamics is an extremely useful computational tool for fluid dynamicists, the complex Hilbert Space is an extremely useful computational tool for quantum physicists. However, we shouldn't take these tools too seriously. If we took inviscid fluid dynamics seriously we would end up believing things that are manifest nonsense e.g. that aeroplanes could never fly. Similarly, if we take Hilbert Space too seriously we will end up believing things that are also manifest nonsense, like half-alive half-dead cats, or particles that go through more than one diffraction slit at the same time! 

The notion of the Hilbert Space as a singular limit of the space of bit strings is central to the re-interpretation of Bell's inequality, discussed below. 

\subsection{Measurement}
\label{planck}

One of the key problems of standard quantum theory is the measurement problem. Conventionally, the evolution of the quantum state during measurement is seen as something additional to Schr\"{o}dinger evolution. However, measurement is an integral aspect of the fractal structure of $I_U$, and hence the physics of the measurement problem is easily described in invariant set theory, as discussed in this Section. This is consistent with the Hilbert Space notion of state superposition as having no fundamental status within invariant set theory. 

As mentioned earlier, the two groupings of intervals in $C^{(2)}_k(t_f)$ are considered as distinct attractors on $I_U$ and correspond to discrete measurement outcomes. As discussed in this section, the existence of such attractors on $I_U$ is consistent with the notion that gravity itself may be a key process acting during quantum measurement (e.g. \cite{Diosi:1989} \cite{Penrose:1994}). Hence the probability conserving dynamical evolution described by $\mathcal D: C(t_i) \rightarrow C(t_f)$, as described above, is consistent with the dynamical evolution of the Cantor Set from a quasi-uniform to a highly lacunar (i.e. `gravitationally clumped') form in state space. 

In Section \ref{complex} state-space trajectory segments were labelled using the symbols `$a$' and `$\lnot a$'. By labelling one segment `$a$' and another `$\lnot a$', then these trajectory segments (each one describing a space time) are to be considered physically distinct.  Consistent with numerical estimates in \cite{Diosi:1989} \cite{Penrose:1994} two space-times $\mathcal M_1$ and $\mathcal M_2$ will be said to be physically indistinct if
\be
\label{pe}
\int_{t_0}^{t} E_G(\mathcal M_1, \mathcal M_2) dt < O(\hbar)
\ee
where $E_G$ denotes the gravitational interaction energy associated with these space times (i.e. the energy needed to move the mass distribution in $\mathcal M_1$ to the mass distribution in $\mathcal M_2$ against the gravitational field in $\mathcal M_2$). This notion is well defined in a Newtonian context, but becomes problematic in a general relativistic context because the Principle of Equivalence prevents a pointwise identification of two distinct space times. In this sense, inequality (\ref{pe}) is a criterion which uses $\hbar$ to define situations where the space-times are sufficiently indistinct that a pointwise identification is possible, i.e. where the Principle of Equivalence breaks down. As discussed in Section \ref{gravtheory}, this may have observable cosmological consequences

\begin{figure}
\centering
\includegraphics[scale=0.4]{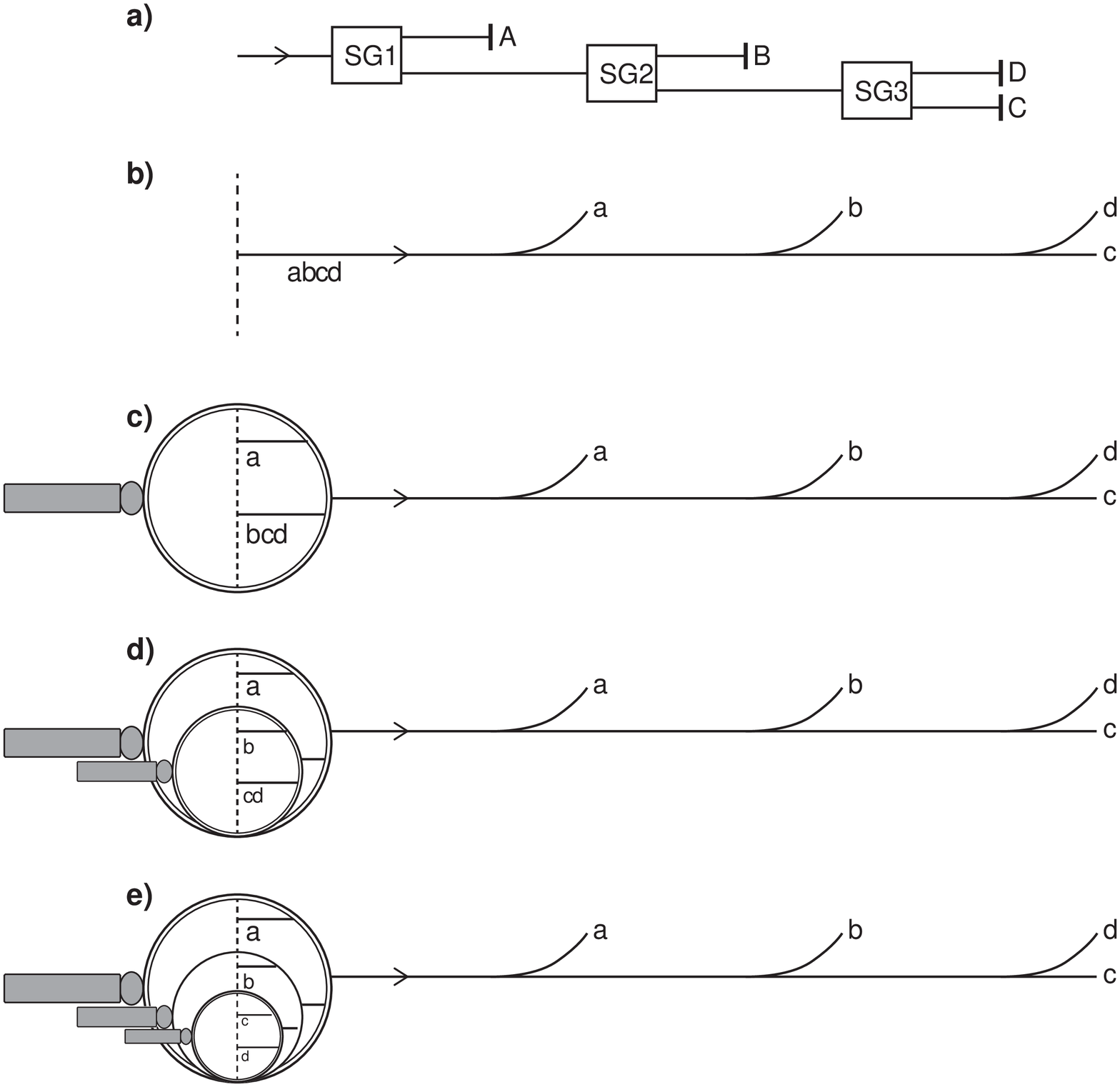}
\caption{a) A sequential Stern-Gerlach experiment. b) State-space trajectories on $C \times \mathbb{R}$ represented at a level of detail where only the intervals of $C_{k_{-1}} (t_0)$ are resolved. c) By zooming into $C$ to reveal the intervals comprising $C_{k_0}(t_0)$, the trajectories labelled `a' can be distinguished from the trajectories labelled `b', `c' and `d'. c) By zooming further into $C$ to reveal the intervals comprising $C_{k_1}(t_0)$, the trajectories labelled 'b' can be distinguished from the trajectories labelled `c' and `d'. d) By zooming further into $C$ to reveal the intervals comprising $C_{k_2}(t_0)$, the trajectories labelled `c' can be distinguished from the trajectories labelled 'd'. In this way, time evolution on $I_U$ is linked to scale on $C$.}
\label{fig:SG}
\end{figure}

To illustrate these concepts (and to illustrate the role of the fractal structure of $C$), consider a spin-1/2 particle fed into a measuring apparatus comprising three sequential Stern-Gerlach (SG) devices (SG1 to SG3: Fig \ref{fig:SG}a). Let us assume the particle is emitted from the source at time $t_0$ and leaves SG1 at time $t_1$. If the particle is deflected `up' by SG1 then it is detected by A at time $t_A$. If the particle is deflected `down' by SG1 and `up' by SG2 then it is detected by B at time $t_B$. Otherwise the particle is detected by either C or D at time $t_{CD}$. 

Let the symbol `$a$' label a space-time in which a particle is emitted by the source at $t_0$ and is deflected `up' by SG1 and detected by A. Similarly, label a space-time in which a particle is emitted by the source and is deflected `down' by SG1 with the symbol `$\lnot a$'. It might be thought that the trajectories associated with these two space-times will have diverged sufficiently to be characterised as distinct by $t_1$. However, this is not the case if (\ref{pe}) is used as the state-space metric characterising distinctiveness. Rather, the divergence between these two space-times is only significant - in the sense that inequality (\ref{pe}) breaks down - when in one of the space-times the particle has entered A and disturbed the motion of the many particles which comprise A, and in the other space-time the particle is deflected `down' by SG1. That is, 
\begin{eqnarray}
\int_{t_{0}}^{t_1} E_G(a, \lnot a) dt < O( \hbar); \;\;\;\;
\int_{t_{0}}^{t_A} E_G(a, \lnot a) dt \nless O(\hbar) 
\end{eqnarray}
Similarly, let the symbol `$b$' label a space-time in which the particle is emitted by the source and is deflected `up' by SG2 and detected by B; `$\lnot b$' labels a space-time in which the particle is emitted by the source and is deflected `down' by SG2. Based on the same estimates, we have 
\begin{eqnarray}
\int_{t_{0}}^{t_A} E_G(b, \lnot b) dt < O(\hbar); \;\;\;\;
\int_{t_{0}}^{t_B} E_G(b, \lnot b) dt \nless O(\hbar) 
\end{eqnarray}
Finally, let `$c$' label a space-time in which the particle is emitted by the source and is deflected `up' by SG3 and detected by C, and `$d$' label a space-time in which the particle is emitted by the source and is deflected `down' by SG3 and detected by $D$. Then 
\begin{eqnarray}
\int_{t_{0}}^{t_B} E_G(c, d) dt < O(\hbar); \;\;\;\;
\int_{t_{0}}^{t_{CD}} E_G(c, d) dt \nless O(\hbar) 
\end{eqnarray}

Fig \ref{fig:SG}b-e illustrates the state-space trajectories associated with these space-times on $C \times \mathbb{R}$ at different levels of approximation.  As shown in Fig \ref{fig:SG}b, an interval of the approximation $C_{k_{-1}}(t_0)$ of $C(t_0)$ is too coarse to be able to distinguish between the `$a$', `$b$', `$c$' and `$d$' trajectories at initial time. Hence an interval is labelled with the single composite `$abcd$' label. At this level the corresponding trajectory appears as a single trajectory which subsequently splits (reminiscent of the many-worlds interpretation, discussed in more detail below!). 

However, by zooming into the finer-scale approximations of $C(t_0)$, the $a$, $b$, $c$ and $d$ trajectories can start to be distinguished and any sense that the trajectories are splitting non-deterministically can be seen to be illusory. In Fig \ref{fig:SG}c we have zoomed in to the $C_{k_0}$ level of $C$. This allows us to distinguish the `$a$'-labelled trajectories from the `$\lnot a$'-labelled trajectories (the latter comprising the `$b$', `$c$' and `$d$' trajectories). In Fig \ref{fig:SG}d we have zoomed in further, to the $C_{k_1}$ level. This allows us to separate the `$\lnot a$' trajectories in Fig \ref{fig:SG}c into those labelled `$b$' and $\lnot b$ (the latter comprising `$c$' and `$d$' trajectories). Finally, in Fig \ref{fig:SG}e we have zoomed in further to $C_{k_2}$. This allows us to separate the trajectories labelled `$\lnot b$' in Fig \ref{fig:SG}d into those labelled `$c$' and `$d$'. 

Now in any particular experiment with such sequential SG devices, the fraction of occurrences that the detectors A, B, C and D register particles depends on the orientation of the devices SG1-SG3. The next stage in the construction is to relate these orientations with the frequency of occurrence of the labels $a$ and $\lnot a$ on $C_{k_0}$, of $b$ and $\lnot b$ on $C_{k_1}$, and of $c$ and $d$ on $C_{k_2}$. As discussed in Section \ref{complex}, these frequencies are functions of the groupings to which the intervals belong, (\ref{pf}), and relate to the parameter $\alpha$ in the quaternion operators $\mathbf{E}^\alpha_\beta$. 

\section{Emergent Nonclassical Properties}
\label{emergent}

Here we are now in a position to discuss the emergence from invariant set theory of one of the key properties of quantum physics: existence of incompatible observables. This will lead directly to the nullification of the Bell Theorem. 

\subsection{Simultaneous Measurements and Incompatible Observables}
\label{simul}

To illustrate the concept of incompatible observables, consider again the standard sequential Stern-Gerlach experiment (Fig \ref{fig:SG}a). We suppose SG1 prepares a spin-1/2 particle with spin is oriented in the $\mathbf{\hat z}$ direction. Experimenter Alice then freely chooses an orientation for SG2 and experimenter Bob freely chooses an orientation for SG3. 

\begin{figure}
\centering
\includegraphics[scale=0.7] {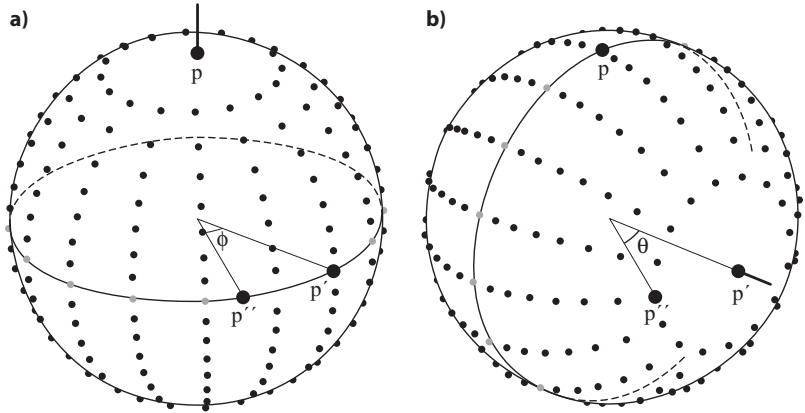}
\caption{The symbolic sequences $E_\beta^{\alpha}(aaa\ldots a)$ represented by `dyadic rational points' on the sphere, relative to two polar orientations $p$ and $p'$. The granularity of the rational points of the sphere make a) and b) incommensurate. For example, using simple number theory (see Appendix C) it can easily be shown that $p''$ is a dyadic rational point in b) but not in a). This is consistent with the notion of incompatible observables in quantum theory, and allows invariant set theory to evade the Bell Theorem, despite being deterministic and locally causal.}
\label{fig:sphere}
\end{figure}

Through the relationships (\ref{incom}), the symbolic sequences $E_\beta^{\alpha}(aaa\ldots a)$ label the iterations $C_k$ of $C$ and define the measurement choices available to Alice and Bob. To understand this better, let us return briefly to the simple `middle thirds' Cantor Set (Fig \ref{fig:cantor}). For any interval of the approximation $C_{k-1}$ are two intervals of $C_{k}$. These intervals can be represented by points on the line associated with the integers 0 and 2. By analogy, the sphere in Fig \ref{fig:sphere}a shows a small collection of the possible dyadic rational points which represent the approximation $C_{k_1}$ in Fig \ref{fig:SG} associated with measuring apparatus SG2. In particular, the North Pole $p$ corresponds to the $\mathbf{\hat z}$ direction. From (\ref{incom}), the North Pole $p$ is associated with the sequence $E_\beta^{0}(aaa\ldots a)= aaa \ldots a$, and the antipodal South Pole with $E_\beta^{2}(aaa\ldots a)= \lnot a \lnot a \lnot a \ldots \lnot a$. More generally, according to the probability function (\ref{pf}), the fraction of `$a$' symbols in the symbolic sequence associated with dyadic rational points at co-latitude $\theta$ is equal to $\cos^2 \theta/2$. Sequences associated with antipodal rational points are always precisely anti-correlated. From (\ref{correspondence}), the spheres in Fig \ref{fig:sphere} correspond to the continuum Bloch sphere of quantum theory.  The rational points of Fig \ref{fig:sphere}a define measurement orientations (dense in the limit $N \rightarrow \infty$) available to Alice.

Suppose for the sake of argument that Alice chooses to orient SG2 in the direction $p'$ in Fig \ref{fig:sphere}a. Then Fig \ref{fig:sphere}b illustrates (some of) the choices available to Bob as he orientates SG3. The set of rational points on the sphere in Fig \ref{fig:sphere}b represents the approximation $C_{k_2}$ in Fig \ref{fig:SG} associated with SG3. Again, in the limit $N \rightarrow \infty$, Bob has a dense set of orientations to choose from. From (\ref{incom}), the North Pole $p'$ is associated with the sequence $E_\beta^{0}(aaa\ldots a)= aaa \ldots a$, and the antipodal South Pole with $E_\beta^{2}(aaa\ldots a)= \lnot a \lnot a \lnot a \ldots \lnot a$. More generally, according to the probability function (\ref{pf}), the fraction of `$a$' symbols in the symbolic sequence associated with dyadic rational points at co-latitude $\theta$ relative to $p'$ is equal to $\cos^2 \theta/2$. For simplicity, suppose Bob chooses to orient SG3 relative to the direction $p''$ shown in Fig \ref{fig:sphere}b. 

Let us now ask the key question: Even though Alice could choose from an effectively dense set of possible orientations, could she nevertheless have oriented SG2 in the direction associated with Bob's choice of $p''$? If she could, then invariant set theory would allow simultaneous measurements in directions associated with both $p'$ and $p''$. This would obviously be inconsistent with quantum physics, rendering invariant set theory dead in the water. An inability to make simultaneous measurements (for incompatible observables) lies at the heart of the uncertainty principle and the non-commutativity of operators in quantum theory. Fortunately for invariant set theory, simple number theory can be used to establish that $p''$ is not available to Alice. To see this, note in Fig \ref{fig:sphere}b that $\cos \theta_{p'p''}$ must be dyadic rational. Let $0<\phi= \theta_{p'p''}< \pi/2$ as in Fig \ref{fig:sphere}a. Then, from the theorem in Appendix C, $\phi$ cannot be a dyadic rational multiple of $\pi$ if $\cos \phi$ is dyadic rational. Since $p''$ is not a dyadic rational point relative to $p$ it is therefore not a possible choice for Alice. In other words, measurements relative to $p'$ and $p''$ are not possible simultaneous measurements. Alice has all the choices in the world - effectively a dense set of orientations on the sphere. But this is not quite the same as saying her choices constitute a continuum. In practice, $N$ is a finite number. However, if the results above hold in the limit $N \rightarrow \infty$, they will certainly hold when $N$ is finite. It is the `granularity' of the rational points, the fact that $\mathbf{E}_{\beta}(aaa \ldots a)$ does not depend continuously on $\beta$ and the nonlinearity of the cosine function, that produces this indeterminacy. 

It is well known that in quantum theory the impossibility of making simultaneous measurements (for non-commuting observables) lies at the heart of Heisenberg's Uncertainty Principle. In this paper, by focussing on qubit physics, it is not possible to derive the traditional `position-momentum' uncertainty relations for non-commuting observables with continuous spectra; to do this would require analysis of systems described by multiple bit strings. This will be discussed elsewhere. However, it is conjectured here that such uncertainty relations are founded on the type of indeterminacy discussed in this Section. It is interesting to note that, in his early papers, Heisenberg refers to the Uncertainty Principle by the expression `Das Unbestimmtheitsprinzip' - the Indeterminacy Principle. From the invariant set perspective, this latter name seems the more appropriate. 

\subsection{The Bell Theorem Revisited}
\label{thebelltheorem}

Let us now return to the Bell Theorem. As discussed in Section \ref{belltheorem}, the measure-zero nature of $I_U$ allows the possibility to nullify the Bell inequality through a partial violation of the measurement independence postulate. This possibility is now discussed in more detail. 

In invariant set theory, a bit string such as $S=\mathbf E^\alpha_\beta(ddd \ldots d)$ can define a sample space of occurrences of disagreement or agreement of pairs of spin measurements on entangled spin-1/2 particles 1 and 2, subject to (\ref{opposite}) and associated with a sample space $\Lambda$ of hidden variables. (A full description of how the operators $\mathbf E^\alpha_\beta$ can be used to describe entanglement will be provided in a paper currently in preparation.) In particular, the label $d$ denotes an occurrence where one measurement is spin `up' and the other spin `down' whilst the label $\lnot d$ denotes an occurrence where either both measurements are `up' or both `down'. The relative angle $\theta_{ab}$ between the two measurement devices with orientations $\mathbf{\hat a}$ and $\mathbf{\hat b}$ is related to $\alpha$ by (\ref{incom}). For $\lambda \in \Lambda$, the set of all relative angles $\theta_{ab}$ for which the triple $(\mathbf{\hat a}, \mathbf{\hat b}, \lambda)$ corresponds to a state of physical reality, is the set where $\cos \theta_{ab}$ is describable by $N$ bits and hence is dyadic rational, c.f. Section \ref{quantum}.

In assessing whether invariant set theory is constrained by the Bell inequality (\ref{bell}), consider three points $a$, $b$ and $c$ on the unit sphere corresponding to the three (measurement) directions $\mathbf{\hat a}$, $\mathbf{\hat b}$ and $\mathbf{\hat c}$ in (\ref{bell}) and defining a spherical triangle $\triangle abc$. If we fix $\mathbf{\hat a}$ as the North Pole of the sphere, then $(\mathbf{\hat a}, \mathbf{\hat b}, \lambda)$ and $(\mathbf{\hat a}, \mathbf{\hat c}, \lambda)$ correspond to states on $I_U$ providing the cosine of the colatitudes of $b$ and $c$ are describable by $N$ bits. So far so good. But for invariant set theory to be constrained by the Bell inequality (\ref{bell}), the third triple $(\mathbf{\hat b}, \mathbf{\hat c}, \lambda)$ must also correspond to a state on $I_U$. This implies that $\cos \theta_{bc}$ must also be dyadic rational. However, by the cosine rule for spherical triangles
\be
\label{triangle}
\cos \theta_{bc}=\cos \theta_{ab} \cos \theta_{ac}+\sin \theta_{ab} \sin \theta_{ac} \cos \phi
\ee
where $\phi$ is the angle subtended by the two great circles of $\triangle abc$ at $a$. Now the first term on the right hand side of ($\ref{triangle}$) is dyadic rational by definition. However, by (\ref{incom}), $\phi/\pi$ must also be dyadic rational and by the discussion in Section \ref{simul} and Appendix C, no $0<\phi< \pi/2$, $\cos \phi$ can be dyadic rational. Hence $\cos \theta_{bc}$ cannot be dyadic rational. (We may wish to consider a physical situation where $a$, $b$ and $c$ are approximately collinear. In this case $\phi$ will be small, but with vanishing probability for large enough $N$ cannot be precisely 0.) Similar arguments can be made with the North Pole at $b$ or $c$. In every case it is impossible for all three correlations in the Bell inequality to be simultaneously defined, no matter how large is $N$. Hence, consistent with a partial violation of the measurement independence postulate,  invariant set theory is not constrained by the Bell inequality, determinism and causality notwithstanding. This argument extends to the CHSH version of the Bell inequality, but this extension will be discussed elsewhere for lack of space. Moreover, the argument here is reminiscent of the finite-precision loophole put forward by Meyer \cite{Meyer:1999} to nullify the Kochen-Specker theorem (indeed, invariant set theory provides a physical model for this loophole). 

Since it is impossible to simultaneously define all three correlations, one might ask how invariant set theory could predict that the Bell inequality will actually be violated experimentally (and where the three correlations are each well defined)? To discuss this point, it is necessary to note that when an experiment is performed to test the Bell (or CHSH) inequality, a sub-experiment has to be performed to estimate each individual correlation. That is to say, in an experimental test of (\ref{bell}), the experimentally correlations can be written
\begin{eqnarray}
Corr_{\Lambda_1}(\mathbf{\hat a}, \mathbf{\hat b})&=&\sum_{\Lambda_1} A(\mathbf{\hat a}, \lambda) B(\mathbf{\hat b}, \lambda)= -\cos \theta_{ab} \nonumber \\
Corr_{\Lambda_2}(\mathbf{\hat a}, \mathbf{\hat c})&=&\sum_{\Lambda_2} A(\mathbf{\hat a}, \lambda) B(\mathbf{\hat c}, \lambda) = -\cos \theta_{ac} \nonumber \\
Corr_{\Lambda_3}(\mathbf{\hat b}, \mathbf{\hat c'})&=&\sum_{\Lambda_3} A(\mathbf{\hat b}, \lambda) B(\mathbf{\hat c'}, \lambda)= -\cos \theta_{bc'} 
\end{eqnarray}
where $\Lambda_1 \ne \Lambda_2 \ne \Lambda_3$ are finite disjoint spaces of hidden variables, each large enough that the correlation estimates are statistically significant. Now, for experiments to estimate these correlations to be physically realisable (ie are associated with points on $I_U$), then all of $\cos \theta_{ab}$, $\cos \theta_{ac}$ and $\cos \theta_{bc'}$ must be dyadic rational. However, crucially (by the discussion above) $c'$ cannot equal  $c$ precisely because we know from the argument above that $\cos \theta_{bc}$ is not dyadic rational. However, because $2\pi/N$ is smaller than the assumed angular resolution of the measuring instruments, it is possible for the experimenter to ensure that $c' \approx c$ to within the finite-precision accuracy of the measuring apparatuses. 

Since $c'$ and $c$ can be made as close as one likes on the sphere with large enough $N$, is it not possible to use `epsilon-delta' arguments to approximate the bit string at $c$ by the bit string at $c'$, the error of the approximation going to zero as $N \rightarrow \infty$? No! The reason (as mentioned above) is that the bit strings do not vary continuously with respect to the phase angle $\phi$, as can be seen from the construction described in Appendix B. The crucial point here is that an experimenter, by measuring correlations with respect to $\theta_{ab}$, $\theta_{ac}$ and $\theta_{bc'}$ might think she has actually measured correlations with respect to $\theta_{ab}$, $\theta_{ac}$ and $\theta_{bc}$. If the bit strings varied continuously with respect to $\phi$ this would be a correct assumption. However, they don't vary continuously and so the assumption is invalid. (Although the bit strings themselves do not vary continously, the probabilities associated with these bit strings do vary continuously. Hence the argument put forward by Mermin \cite{Mermin} against Meyer's finite-precision construction \cite{Meyer:1999} is not valid here.) 

Whilst superdeterministic theories are counterfactually void, invariant set theory is instead merely counterfactually incomplete. It is on the one hand complete enough to allow a natural definition of probability and correlation, but is on the other hand incomplete enough to allow a partial violation of the measurement independence postulate sufficient to nullify the Bell inequality. This notion of counterfactual incompleteness can be used to resolve other famous conceptual conundrums in quantum physics. Consider the two-slit experiment. The observed diffraction structure when particles are allowed to travel through either slit without detection seems logically inconsistent with the realistic notion that these same particles must have travelled either through one slit or the other. If the slits are far enough apart, this seems causally inconsistent too. However, in an actual situation where both slits are open and a \emph{particular} particle (labelled by $\lambda$) travels though one of the slits $(\mathsf{open}, \mathsf{open}, \lambda)$, a counterfactual world in which \emph{that same particle} travelled through the same slit, but the other slit was closed off $(\mathsf{open}, \mathsf{closed}, \lambda)$, would lie off $I_U$ and hence would not correspond to a state of physical reality. The actual proximity of one slit to the other is irrelevant to this argument. Hence the cosmological invariant set postulate allows one of the most iconic of quantum observations to be explained in a perfectly causal and realistic way.

\section{A Gravitational Theory of the Quantum?}
\label{gravtheory}

If invariant set theory is correct, then quantum theory should not be considered a fundamental theory of physics, but rather is the singular limit of a deeper theory based on state-space geometry. Five lines of evidence from this study suggest that this deeper theory is fundamentally gravitational in character. 
\begin{itemize}
\item $I_U$ is presumed to be the invariant set of a self-contained self-gravitating system: the universe. 
\item Like general relativity, invariant set theory assumes the primacy of geometry in determining dynamical evolution. 
\item As discussed in Section \ref{planck}, the symbolic representation of the invariant set $I_U$ is consistent with the notion of gravitational distinctness.
\item As discussed in Section \ref{universe}, the zero-volume nature of $I_U$ has been linked to state-space convergence at space-time singularities (with the corresponding cosmological-scale forcing associated with the positive cosmological constant). 
\item As discussed in Section \ref{complex}, the determination of the energy (=$\hbar \omega$) of a particle from the divergence of neighbouring trajectories on $I_U$ is analogous to the way in which the kinetic energy of a test particle is determined by the equation for geodesic deviation in general relativity. 
\end{itemize}
This in turn suggests that conventional approaches to the unification of gravity with the other forces of nature are misguided. These standard approaches take as fundamental the axioms (and more particularly the Ans\"{a}tze) of quantum field theory, applying them to some putative lagrangian which includes gravity (cf the second sentence of the Penrose quote in Section \ref{introduction}). However, based on the analysis presented here, this approach may be putting the cart before the horse - the cart being quantum theory and the horse gravitation theory. Rather than trying to formulate a `quantum theory of gravity', perhaps we should instead, as invariant set theory attempts, be trying to formulate a `gravitational theory of the quantum'. Below we discuss three consequences of this \emph{volte face}: black hole information loss, vacuum energy and detection of gravitons. 

There are some profound enigmas reconciling black holes, information and the foundations of physics, which some consider to lie at the heart of the difficulty in formulating a consistent theory of quantum gravity. Indeed this problem has been described as threatening to overthrow the current foundations of physics, and as posing a crisis with similarities to the classical crises which led to the development of quantum theory itself \cite{Giddings}. Invariant set theory provides a novel solution to this foundational crisis. 

The notion of information, like entropy, is not a primary concept in standard theoretical physics, but rather derives from Boltzmann's concept of coarse-grained variations of microstates defined over volumes in state space (i.e. the full-volume Euclidean space in which the measure-zero invariant set $I_U$ is embedded). Penrose \cite{Penrose:2004} \cite{Penrose:2010} argues that black-hole information loss can be interpreted in terms of state-space convergence (in this full volume Euclidean space) associated with space-time singularities. Others argue that if unitarity is violated in black-holes, it will be violated everywhere \cite{Susskind:2008} - conservation of probability can never be guaranteed. This is one aspect of the `crisis' referred to above. By contrast, as discussed in Section \ref{quantum}, conservation of probability is not violated by evolution on $I_U$. Rather it is violated by mathematical transformations which take states on $I_U$ to physically unrealisable counterfactual states in the embedding space off $I_U$. This is relevant to the discussion about information loss.  More specifically, there can be no physical information loss in invariant set theory because states of physical reality are undefined in the embedding Euclidean space off $I_U$. Put another way, the notion of information loss in the (full-volume) Boltzmannian sense can exist peaceably with the cosmological invariant set postulate where there is no physical information loss. The argument here is quite similar to that for the Bell Theorem. As discussed in Section \ref{thebelltheorem}, invariant set theory is not constrained by Bell's inequality because Bell's inequality necessarily requires us to consider states off the invariant set, and in invariant set theory, these states are undefined. Hence the Bell inequality defined by states off the invariant set can exist peaceably with local causality (and determinism) on the invariant set. This whole discussion can be broadened to include the relationship between invariant set theory and the second law of thermodynamics; this will be done elsewhere. In short, whilst the black-hole information paradox may well pose a crisis for conventional approaches to quantum gravity, it is not problematic for invariant set theory where geometry and hence gravity are primary concepts. 

Let us now briefly mention another consequence of invariant set theory The $>100$ orders of magnitude discrepancy between quantum field theoretic estimates of the cosmological constant and the observed value is well known. Zero-point vacuum fluctuations are constrained by the Heisenberg energy-time uncertainty principle. According to the discussion in Section \ref{planck}, two space-times $\mathcal M_1$ and $\mathcal M_2$ which differ solely in terms of such miniscule vacuum fluctuations would be deemed indistinct according to criterion (\ref{pe}). As such, in invariant set theory, such minimal fluctuations in space time would not be gravitationally coupled and would therefore not contribute to the cosmological constant. Rather, as mentioned in Section \ref{universe}, large-scale forcing can be associated with the repulsive cosmological constant, whilst small-scale irreversibility can be associated with Planck-scale state-space trajectory convergence. Just as with other multi-scale systems (such as the turbulent Navier-Stokes equations), these large-scale and small-scale forcings must be in some overall balance in order that $I_U$ is structurally stable. For example, too much state space convergence (or too weak large-scale forcing) and $I_U$ could not be fractal, but would rather collapse to a limit cycle or fixed point. Too much large-scale forcing (or too weak small-scale convergence) and $U$ would not be confined to a compact set in state space. Since the Planck-scale state-space trajectory convergence is largely determined by the amount of dark matter in the universe, then the existence of a structurally stable fractal $I_U$ (from which quantum physics can be emergent) implies some close balance between dark matter and dark energy, of a type not hitherto considered. The possible insights that invariant set theory may provide into the dark universe will be explored elsewhere. 

Finally, we might consider the status of the graviton: the quantum excitation of the gravitational field as would be predicted by standard quantum field theoretic axioms. Particle and gravitational wave detectors are much too insensitive to be able to detect individual gravitons and this will be the case for the foreseeable future. However, since invariant set theory rejects as fundamental standard field quantisation procedures, then the theory must similarly reject the notion of the graviton. That is to say, it is conjectured here that an inability to detect gravitons is not fundamentally because of detector insensitivity, but rather because there is no such thing as a graviton. In this respect it can be noted that Freeman Dyson \cite{Dyson:2013} has noted that the rather general Bohr-Rosenfeld argument for the quantisation of the electromagnetic field cannot be applied to the gravitational field - there is no compelling case within standard physics that the gravitational field must be subject to conventional field quantisation. 

\section{Relationships to Other Approaches to Quantum Physics}
\label{others}

How does invariant set theory compare with other models of quantum physics? These days, a standard model for quantum measurement is provided by the notion of decoherence, where state evolution is strictly unitary at all times (cf the first two columns of Table 2). Consistent with this, the notion of a superposed state is fundamental and irreducible. An alternative to standard quantum theory is provided by appending collapse models to Schr\"{o}dinger evolution during the measurement process \cite{Pearle:1976}. Hence, in these models, unitary evolution breaks down during the measurement process. A key motivation for such models is that after measurement, the system is no longer in a superposed state. Although time evolution is not unitary in a quantum theory with explicit collapse, all counterfactual transformations of the type discussed above continue to be describable by unitary operators acting on the quantum state vector (cf third and fourth columns of Table 2). Often, these collapse models (e.g. the `Objective Reduction' model of Penrose \cite{Penrose:1994}) have their physical basis in gravitation theory. 

Invariant set theory provides a `third way'. Like quantum theory with collapse, invariant set theory assumes an objectively realistic macroscopic world. However, invariant set theory is realistic on the microscale too. In invariant set theory, the notion of unitarity is defined through the quaternion operators $\mathbf{E}_\beta$. For example, as shown in Appendix B, $\mathbf{E}_\beta$ can be represented as unitary matrices, i.e. where $\mathbf{E}_\beta \mathbf{E}^*_\beta=\textrm{id}$ and $\mathbf{E}^*_\beta$ is the matrix transpose of $\mathbf{E}_\beta$ with all occurrences of the permutation/negation operator $i$ replaced by $-i$. In invariant set theory, as in decoherent quantum theory, real-world evolution from preparation to measurement is described by these unitary operators. On the other hand, as we have discussed above, unitarity breaks down in invariant set theory in considering transformations to certain unphysical counterfactual worlds off the invariant set (see fifth and sixth columns of Table 2). 

\vspace{7 mm}
\begin{center}
\begin{tabular}{|c|c|c|} \hline
quantum theory                           & time evolution & unitary \\ \cline{2-3}
with decoherence                & counterfactual $\textrm{X}^n$& unitary \\ \hline
quantum theory                & time evolution & non-unitary \\ \cline{2-3}
with collapse                     & counterfactual $\textrm{X}^n$& unitary \\ \hline
invariant set          & time evolution & unitary \\ \cline{2-3}
theory                                         & counterfactual $\textrm{X}^n$& non-unitary \\ \hline
\end{tabular}
\\
\vspace{2mm}
TABLE 2
\end{center} 

It is worth commenting on relationships between invariant set theory and yet other approaches to quantum theory. For example, the de Broglie-Bohm interpretation is also a deterministic approach to quantum physics \cite{BohmHiley}, but is explicitly non-local. This non-locality is manifest in the so-called quantum potential $Q$ acting in configuration space. Relative to invariant set theory, the quantum potential can be considered an approximate `smoothed-out' and hence differentiable representation in configuration space of the highly non-differentiable invariant measure $\mu$ of $I_U$ in the full state space. (One could similarly imagine a differentiable potential function approximating $I_L$ in the reduced $X-Y$ subspace of the Lorenz state space). Because of its smoothness, the support of $Q$ has full measure in configuration space and the resulting probability distributions loose the property which allows the theory to be unconstrained by the Bell inequality. As a result, de Broglie-Bohm theory cannot be locally causal. 

This raises the following question: If the measure of $I_U$ is too singular to be describable by the tools of differential (eg Riemannian) geometry, with what type of geometry can $I_U$ and its invariant measure be described? The answer is noncommutative geometry \cite{Connes} \cite{Lapidus}; one of the applications of noncommutative geometry is to define rigorously singular fractal measures. This suggests that there may be profound links - to be discovered - between invariant set theory and the programme to reformulate physics using the abstract formalism of non-commutative geometry and algebra. 

Finally, let us consider the link between invariant set theory and the many-worlds interpretation of quantum theory. Fig \ref{fig:SG}b is suggestive of the `world-splitting' characteristic of the many-worlds interpretation. However, this is illusory. As shown in Fig \ref{fig:SG}c-e, there is no such splitting. Rather, trajectories which are close together at initial time, diverge exponentially. In addition, these neighbouring trajectories should not be thought of as representing alternative `parallel' universes to our own.  Rather, in invariant set theory, each neighbouring trajectory should be considered part of our own universe, albeit associated with some remote future aeon. As mentioned above, the Bohmian notion of implicate and explicate order is relevant in capturing this concept, i.e. of space-time segments being neighbours in the implicate order associated with $I_U$, but distant in the explicate order associated with time. 

\section{Conclusions}
\label{conclusions}

In this paper a physical postulate has been introduced which provides some new perspectives firstly on determinism and causality in fundamental physics, and secondly on the programme to unify gravitation and quantum physics. The `cosmological invariant set postulate' states that the universe $U$ is evolving on a measure-zero fractal invariant set $I_U$ in its state space. It is motivated by recent advances in observational and theoretical cosmology, and by the state space geometry of certain nonlinear dynamical systems. 

To summarise what has been achieved, let us return to Fig \ref{fig:triangle}. Least importantly perhaps, the analysis in this paper makes clear that the notion of chaos can be formulated in a perfectly relativistic manner (bottom edge of triangle). In particular, a relativistic definition of chaos should be based on geometric properties of the corresponding invariant set, such as its fractional dimension, rather than non relativistic concepts such as Lyapunov exponent (see also \cite{Cornish:1997}). More importantly, by constructing a symbolic representation of $I_U$ using permutation/negation representations of complex numbers, we can account for many of the conceptual conundrums of quantum theory within a purely deterministic and locally causal framework (right hand edge of triangle). In particular, the experimental violation of Bell inequalities is ascribable to the non-trivial measure-zero structure of the perfectly causal set $I_U$ in state space, and not to any breakdown of determinism or local causality in space-time. The analysis has suggested that the complex Hilbert Space of quantum theory can be treated as the singular limit of probabilistic descriptions of the symbolic representation of $I_U$. Most importantly, perhaps, this has suggested a new approach to the unification of gravitation theory and quantum theory, where the quantum cart is not put before the gravitational horse (left hand edge of triangle). This has led to novel speculations about the nature of the dark universe and to the physical status of the graviton.  

Of course, there is still much to be achieved. The discussion in this paper has been entirely in terms of individual and entangled qubits. The extension of invariant set theory to observables with continuous spectra, hence position and momentum, is currently being development. In addition, although we have discussed how differences between the classical Liouville equation and the Schr\"{o}dinger equation ($i$, $\hbar$ and the Hilbert Space) can be explained in a causal and deterministic framework, a specific `invariant set' alternative to the Schr\"{o}dinger equation (or the relativistic Dirac equation) has not been proposed. A key variable in such an equation is the invariant measure $\mu$ of $I_U$. However, this immediately raises the question of what mathematical framework would be needed to describe an evolution equation based on such a singular variable. For example, $\mu$ is much too irregular to be treated using the conventional calculus. As discussed above, the mathematics of non-commutative geometry may be needed, not because fractal geometry is itself non-commutative, but because the mathematical structures which arise in non-commutative geometry (e.g. spectral triples) are powerful enough to also treat commutative singular spaces of the type discussed here \cite{Connes} \cite{Lapidus} \cite{Guido}. Indeed, based on this, one can begin to understand how non-singular but non-commutative variables - generic quantum observables - arise naturally in a description of the singular but commutative properties of $\mu$.  

In conclusion, motivated by Ed Lorenz's great contributions to the field of nonlinear dynamics and the physical insights of Roger Penrose concerning the nature of quantum physics and relativistic cosmology, the elements of a theory have been developed, based not on the calculus of fields in space-time, but instead on measure-zero geometry in state space. This geometry links to some of the deepest theories of 20th Century mathematics, not least G\"{o}del's incompleteness theorem. Invariant set theory attempts to extend Einstein's great insight about the primacy of geometry in fundamental physics. Perhaps fittingly, this theory not only strongly supports Einstein's beliefs about the deterministic causal nature of reality (no dice, no spooky action), it also suggest that the quantum world is profoundly subservient to the geometric phenomenon we call gravity, and without incorporating this into out theories of fundamental physics, we will never satisfactorily unify gravity and quantum theory. 

\section*{Acknowledgements}

My thanks to Dr Irene Moroz for Fig \ref{fig:upo}, and to Drs Andreas D\"{o}ring, Lucien Hardy, Adrian Kent, Terry Rudolph, Antje Weisheimer and anonymous referees for very useful discussions and correspondence.

\bibliography{mybibliography}

\appendix

\section{The Cantor Set $C^{(N)}$}

Let $C^{(N)}_{k}$ comprise $2^N(2^N+1)$ copies of $C^{(N)}_{k-1}$, each copy reduced by a factor $1/[2^{2N}(2^N+1)]$.  The fractal (similarity) dimension of $C^{(N)}$ is therefore equal to $\log [2^N(2^N+1)]/ \log [2^{2N}(2^N+1)] \approx 2/3$ for large $N$. Consider two realisation of $C^{(N)}$. In particular, $C^{(N)}_k(t_i)$ comprises $2^{N}+1$ uniformly spaced groupings of $2^N$ intervals.  If an interval of $C^{(N)}_{k-1}$ has unit length, the width of a grouping of $C_k^{(N)}(t_i)$ is $1/[2^{N-1}(2^N+1)]$ i.e. $O(2^{-2N})$, whilst the gap between groupings has length $1/2^N-1/2^{N-1}$ i.e.  $O(2^{-N})$. Hence, as illustrated in Table 2, for large $N$ $C_k^{(N)}(t_i)$ has the superficial appearance of a dense set of uniformly-spaced `points' (though on sufficient `zooming', all of these `points' can be seen to be groupings of intervals that contain further structure). By contrast, the highly lacunar $C^{(N)}_k(t_f)$ always comprises just 2 groupings, each of $2^{N-1}(2^N+1)$ intervals. Whilst the width of a grouping of $C_k^{(N)}(t_f)$ is $1/2^N$, the gap between the two groupings has length $1-1/2^{N-1}$, i.e. $O(1)$.  Hence for large $N$, $C_k^{(N)}(t_f)$ appears superficially to comprise just two `points'. 

\vspace{7 mm}
\begin{center}
\begin{tabular}{|c|c|c|c|c|c|} \hline
N & \# Groupings & Width of Grouping & Smallest $\alpha$& \# Groupings & Width of Grouping \\
    & $C^{(N)}(t_i)$ & $C^{(N)}(t_i)$ & in $\mathbf{E}_\beta^\alpha$ & $C^{(N)}(t_f)$ & $C^{(N)}(t_f)$ \\ \hline
  2&5 &$10^{-1}$&1&2 & 1/4 \\
  4&17&$7 \times 10^{-3}$&1/4&2& $1/64$ \\
 64&$\sim 10^{19}$& $\sim 10^{-38}$&$\sim 10^{-19}$& 2& $\sim 10^{-19}$ \\ 
 1024&$\sim 10^{308}$& $\sim 10^{-616}$&$\sim 10^{-308}$& 2& $\sim10^{-308}$ \\ 
 \hline
\end{tabular}
\\
\vspace{2mm}
TABLE 2
\end{center} 

\section{Quaternionic Permutation/Negation Operators} 

The quaternions $\mathbf{E}_0$,  $\mathbf{E}_1$,  $\mathbf{E}_2$ defined in Section \ref{complex} (see (\ref{quaternion})) can be written as  $2^2 \times 2^2$ matrices
\be
\label{three}
\mathbf{E}_0=\bp \;&i\\i&\;\ep;
\;\;\mathbf{E}_1=\bp i&\;\\\;&-i\ep;
\;\;\mathbf{E}_2=\bp \;& 1\\ -1 &\; \ep \nonumber
\ee   
where blank entries denote the zero matrix and where 
\be
i=  \bp \; & 1 \\ \lnot & \; \ep \nonumber
 \ee
Using the notion of self-similarity, the operators $\{\mathbf{E}_0, \mathbf{E}_1\}$ can in turn be used as block matrix elements to generate the four $2^3 \times 2^3$ square-root-of-minus-one operators. 
\begin{align}
\mathbf{E}_{00}=\bp \; &\mathbf{E}_0 \\ \mathbf{E}_0& \; \ep;
\;\;\mathbf{E}_{01}&=\bp \; & \mathbf{E}_1 \\ \mathbf{E}_1& \; \ep \nonumber \\
\mathbf{E}_{10}=\bp \mathbf{E}_0 & \;\\ \; & -\mathbf{E}_0 \ep;
\;\;\mathbf{E}_{11}&=\bp \mathbf{E}_1 & \;\\ \; & -\mathbf{E}_1 \ep \nonumber
\end{align}
which satisfy the following quaternionic relationships:
\begin{align}
\label{morequats}
\mathbf{E}_{00} \circ \mathbf{E}_{10} = 
\mathbf{E}_{01} \circ \mathbf{E}_{11} = \bp \;& \mathbf{1}\\ \mathbf{-1} &\; \ep \nonumber
\end{align}
These permutation/negation operators can be ordered in the set
\be
\{\mathbf{E}_{00}, \mathbf{E}_{01}, \mathbf{E}_{10}, \mathbf{E}_{11} \} \nonumber
\ee
and by self similarity can again be used to generate the $2^4 \times 2^4$ matrices
\begin{align}
\label{fifteen}
\mathbf{E}_{000}=\bp \; &\mathbf{E}_{00} \\ \mathbf{E}_{00}& \; \ep;  
\;\;\mathbf{E}_{001}=&\bp \;& \mathbf{E}_{01} \\  \mathbf{E}_{01} & \; \ep;
\dots
\;\;\mathbf{E}_{011}=\bp \; & \mathbf{E}_{11} \\ \mathbf{E}_{11} & \; \ep \nonumber \\
\mathbf{E}_{100}= \bp \mathbf{E}_{00} & \;\\ \; & -\mathbf{E}_{00} \ep;
\;\;\mathbf{E}_{101}=&\bp \mathbf{E}_{01} & \; \\ \;& -\mathbf{E}_{01} \ep;
\dots
\mathbf{E}_{111}=\bp  \mathbf{E}_{11} &\; \\ \;& -\mathbf{E}_{11} \ep \nonumber
\end{align}
which satisfy quaternionic relationships
\be
\mathbf{E}_{000} \circ \mathbf{E}_{100} = \mathbf{E}_{001} \circ \mathbf{E}_{101}= \ldots = \mathbf{E}_{011} \circ \mathbf{E}_{111}=\bp \;& \mathbf{1}\\ \mathbf{-1} &\; \ep \nonumber
\ee
In turn, these operators form an ordered set
\be
\label{seq}
\{\mathbf{E}_{000}, \mathbf{E}_{001}, \mathbf{E}_{010}, \mathbf{E}_{011}, \mathbf{E}_{100}, \mathbf{E}_{101}, \mathbf{E}_{110}, \mathbf{E}_{111}\} \nonumber
\ee
Continuing in this way, sets of independent quaternions can be generated, associated with square matrices whose order equals $N$, where $N$ is any power of 2. If we insert a radix point after the first digit in each of the subscript sequences above, then the ordered set of independent quaternion operators can be written as $\{\mathbf{E}_{\beta}\}$ where $0 \le \beta< 2$ is a dyadic rational number which can be described by $N-1$ bits. If we include the negations $-\mathbf{E}_\beta$ in  $\{\mathbf{E}_{\beta}\}$, then $0 \le \beta < 4$ is a dyadic rational which can be described by $N$ bits. 

Let $\mathbf{\bar E}_\beta$ denote $2^N \times 2^N$ block diagonal matrix containing $2^N/N$ copies of $\mathbf{E}_\beta$. Repeatedly using the fact that for any matrix $A$,  
\be
\bp A & \; \\ \; & A \ep =
\bp \; & 1 \\ A \; \ep \bp \; & 1 \\ A \; \ep  \nonumber
\ee
we can define $\mathbf{\bar{E}}_{\beta}^{\alpha}$ where $\alpha$ is a dyadic rational number which, like $\beta$, can be described by $N$ bits. See \cite{Palmer:2012} for details. The overbar is dropped in the main text above. 

\section{When does the cosine of a rational angle give a rational number?}

$\mathbf{Theorem}$\cite{Jahnel:2005}.  Let $0 < \phi < \pi/2$ and $\phi/\pi \in \mathbb{Q}_2$, the space of dyadic rationals. Then $\cos \phi \notin \mathbb{Q}$. 

We derive a \emph{reductio ad absurdum}. Assume that $\cos \phi = a/b$ is rational, where $a, b \in \mathbb{Z}, b \ne 0$ have no common factors.  Using the identity $2 \cos 2\phi = (2 \cos \phi)^2-2$ we have
\be
2\cos 2\phi = \frac{a^2-2b^2}{b^2}
\ee
Now $a^2-2b^2$ and $b^2$ have no common factors, since if $p$ were a prime number dividing both, then $p|b^2 \implies p|b$ and $p|(a^2-2b^2) \implies p|a$, a contradiction. Hence if $b \ne \pm1$, then the denominators in $2 \cos \phi, 2 \cos 2\phi, 2 \cos 4\phi, 2 \cos 8\phi \dots$ get bigger and bigger without limit. On the other hand, with $0 < \phi/\pi < 1/2 \in \mathbb{Q}$, then $\phi/\pi=m/n$ where $m, n \in \mathbb{Z}$ have no common factors. This implies that the sequence $(2\cos 2^k \phi)_{k \in \mathbb{N}}$ admits at most $n$ values. Hence we have a contradiction. Hence $b=\pm 1$ and $\cos \phi =0, \pm1/2, \pm1$. No $0 < \phi < \pi/2$ with $\phi/\pi \in \mathbb{Q}_2$ has $\cos \phi$ with these values. 

\end{document}